\documentstyle[12pt]{article}

\topmargin - 20mm
\begin{document}

\title{\bf Ising Model and ${\bf Z}_{2}$ Electrodynamics}

\author{Yury M. Zinoviev\thanks{This work was supported in part by the Russian Foundation for Basic
Research (Grant No. 12-01-00094), the Program for Supporting Leading Scientific Schools
(Grant No. 4612.2012.1) and the RAS Program "Fundamental Problems of Nonlinear Mechanics."}}

\date{}
\maketitle

Steklov Mathematical Institute, Gubkin Street 8, 119991, Moscow,
Russia,

 e - mail: zinoviev@mi.ras.ru

\vskip 1cm

\noindent {\bf Abstract.} The correlation functions and spontaneous
magnetization are calculated for the three-dimensional Ising model
and for the three-dimensional ${\bf Z}_{2}$ electrodynamics.

\vskip 1cm

\section{Introduction}

We consider first the two-dimensional Ising model for a square lattice of 
$M$ columns and $N$ rows. The Hamiltonian is
\begin{equation}
\label{1.15}
\overline{H} (\sigma_{11},..., \sigma_{NM}) = - H\sum_{m,n} \sigma_{nm} -
J_{1}\sum_{m,n} \sigma_{nm} \sigma_{n + 1,m} - J_{2}\sum_{m,n} \sigma_{nm}
\sigma_{n,m + 1}.
\end{equation}
$\sigma_{nm}$ is a classical variable taking on the values $\pm 1$.
For boundary conditions we can assume either that the
lattice is wrapped on a torus or we can assume that the lattice has
free ends. For an $M\times N$  lattice the free energy per site and
the magnetic moment per site (magnetization) are defined by
$$
F_{MN}(H) = \frac{1}{\beta MN} \ln \left( \sum_{\sigma_{nm} \, =\, \pm 1}
\exp \{ - \beta \overline{H} (\sigma_{11},..., \sigma_{NM}) \} \right),
$$
\begin{equation}
\label{1.13} M_{MN}(H) = \frac{1}{MN} \frac{\sum_{\sigma_{nm} \, =\, \pm 1} \sum_{m,n}
\sigma_{mn} \exp \{ - \beta \overline{H} (\sigma_{11},..., \sigma_{NM})
\}}{\sum_{\sigma_{nm} \, =\, \pm 1} \exp \{ - \beta \overline{H} (\sigma_{11},..., \sigma_{NM}) \}} =
\frac{\partial F_{MN}}{\partial H}.
\end{equation}
The constant $\beta = (kT)^{- 1}$ is positive. Yang \cite{3} calculated the
spontaneous magnetization
\begin{equation}
\label{1.14} M_{Yang} = |\lim_{\alpha \rightarrow 0+} \lim_{M,N
\rightarrow \infty} M_{MN}(\alpha /M)|.
\end{equation}
$M$ and $N$ tend to infinity together, i.e., with $M/N$ a fixed ratio.
The spontaneous magnetization of Montroll, Potts and Ward \cite{5} is
\begin{equation}
\label{1.16} M_{MPW}^{2} = \lim_{M,N \rightarrow \infty} (MN)^{- 2}
\sum_{n,m,n^{\prime},m^{\prime}} <\sigma_{n^{\prime}m^{\prime}}^{x}
\sigma_{nm}^{x} >_{MN},
\end{equation}
where $<\cdots >_{MN}$ denotes a thermal average in zero field for
an $M\times N$ lattice. In order to describe the situation in the 
two-dimensional Ising model we cite the paper \cite{1}:

\noindent "Although a great deal of effort has been spent on the
two-dimensional Ising model, the amount of exact results is
remarkably limited. For the case of the rectangular lattice without
magnetic field Onsager and Kaufman \cite{2} - \cite{14} have given
the free energy per lattice site and also the correlation functions
for spins at finite distances. In particular, it is readily observed
that the expression for the two-spin correlation function becomes
rapidly more and more complicated as the separation between the two
spins increases. It is for this reason that it is quite difficult to
obtain, as first accomplished by Yang \cite{3}, the spontaneous
magnetization, which is closely related to the limiting value at
infinite separations of the two-spin correlation function."

Schultz, Mattis and Lieb \cite{6} write on the paper of
Yang \cite{3} and on the paper of Montroll, Potts and Ward \cite{5}:

"In contrast to the free energy, the spontaneous magnetization of
the Ising model on a square lattice, correctly defined, has never
been solved with complete mathematical rigor. Starting from the only
sensible definition of the spontaneous magnetization, the methods of
Yang, and of Montroll, Potts, and Ward are each forced to make an
assumption that has not been rigorously justified. The assumptions
appear to be quite different; however, from the similarities between
the difficulties encountered in trying to justify them, and the
identity of the results obtained, one might conclude that they are
closely related."

Isakov \cite{7} obtained the estimates of the derivatives of the
magnetization with respect to the magnetic field for zero magnetic
field and sufficiently large $\beta J_{1}$ and $\beta J_{2}$. These
estimates imply that the magnetization cannot be a holomorphic
function of the magnetic field.

In the paper \cite{9} the new definition of the spontaneous
magnetization was suggested by making use of the one-dimensional
Ising model. Let the number $\sigma_{k} = \pm 1$, $k = 1,...,N + 1$,
$\sigma_{N + 1} = \sigma_{1}$, be given. The partition function of
the Ising model with the constant magnetic field $H(k) = H$,
$k = 1,...,N$, is
\begin{equation}
\label{1.2} Z_{0}(J,H;T(1,N)) = \sum_{{\sigma_{k} \, =\, \pm \, 1,\,
\, k\, =\, 1,...,N + 1,} \atop {\sigma_{N + 1} \, =\, \sigma_{1}} }
\exp \{ \beta J\sum_{k\, =\, 1}^{N}
\sigma_{k} \sigma_{k + 1} +
\beta H \sum_{k\, =\, 1}^{N} \sigma_{k} \}.
\end{equation}
Due to (\cite{8}, Chapter II, formula (4.5), Chapter III, formulas (2.10),
(2.13)) the partition function and  the average total magnetization are
\begin{eqnarray}
\label{1.3}
Z_{0}(J,H;T(1,N)) = (\lambda_{+} (J,H))^{N} + (\lambda_{-} (J,H))^{N},
\nonumber \\
\lambda_{\pm} (J,H) = \exp \{ \beta J\} (\cosh \beta H \pm
(\sinh^{2} \beta H + \exp \{ - 4\beta J\} )^{1/2}),
\end{eqnarray}
\begin{equation}
\label{1.4} \overline{M}_{0} (J,H;T(1,N)) = \beta^{- 1}
\frac{\partial}{\partial H} \ln Z_{0}(J,H;T(1,N)),
\end{equation}
\begin{equation}
\label{1.5} \lim_{N \rightarrow \infty} N^{- 1}\overline{M}_{0}
(J,H;T(1,N)) =  (\sinh^{2} \beta H + \exp \{ - 4\beta J\} )^{- 1/2}\sinh \beta H.
\end{equation}
For the vacuum ($J = 0$) the definition (\ref{1.2}) implies
\begin{equation}
\label{1.7} Z_{0}(0,H;T(1,N)) = (2\cosh \beta H)^{N}
\end{equation}
and the magnetization (\ref{1.5}) for $J = 0$ is equal to $\tanh \beta H$.
It seams natural that the magnetization (\ref{1.5}) (magnetic moment per
site) of the vacuum ($J = 0$) should be zero.

One edge has two boundary vertices. The "energy" of the magnetic field for an
edge is the product of the magnetic fields corresponding to the boundary
vertices of edge. Summing up the "energies" of the magnetic field
$H(k) = H, k = 1,...,N$, over all edges we get
\begin{equation}
\label{1.6} \sum_{k\, =\, 1}^{N} H(k)H(k + 1) = NH^{2},\, \, H(N + 1) =
H(1).
\end{equation}
The average total magnetization (\ref{1.4}) and the "energy" (\ref{1.6})
become infinite for $N \rightarrow \infty$. In order to obtain the finite
values in the quantum field theory the "re-normalized" constants are used.
In the definition (\ref{1.14}) Yang used the "re-normalized" constant
magnetic field $H/M = (M/N)^{- 1/2} (MN)^{- 1/2}H$. The lattice has $M$
columns and $N$ rows with the fixed ratio $M/N$. We consider the
"re-normalized" constant magnetic field
\begin{equation}
\label{1.1}
H(k) = N^{- 1/2}\beta^{- 1}\tanh \beta H,\, \, k = 1,...,N,
\end{equation}
to get the finite "energy" (\ref{1.6}) for $N \rightarrow \infty$.
In view of the relations (\ref{1.3}), (\ref{1.7}) we get the spontaneous
magnetization for the "re-normalized" constant magnetic field (\ref{1.1})
\begin{equation}
\label{1.8} \lim_{N \rightarrow \infty} \frac{\partial}{\partial x}
\left( \ln \frac{Z_{0}(J,N^{- 1/2}\beta^{- 1}x;T(1,N))}{Z_{0}(0,N^{-
1/2}\beta^{- 1}x;T(1,N))} \right)_{x\, =\, \tanh \beta H} = (\exp \{
2\beta J\} - 1)\tanh \beta H.
\end{equation}
For the vacuum ($J = 0$) the spontaneous magnetization (\ref{1.8}) is
equal to zero. The value (\ref{1.8}) is called the spontaneous
magnetization since the "re-normalized" constant magnetic field (\ref{1.1})
tends to zero when $N \rightarrow \infty$.

Due to (\cite{8}, Chapter III, formula (3.1)) the two-spin
correlation function
\begin{equation}
\label{1.9} < \sigma_{m} \sigma_{n} >_{N} \, =
(Z_{0}(J,0;T(1,N)))^{- 1} \sum_{{\sigma_{k} \, =\, \pm \, 1,\,
\, k\, =\, 1,...,N + 1,} \atop {\sigma_{N + 1} \, =\, \sigma_{1}} }
\sigma_{m} \sigma_{n} \exp \{ \beta J\sum_{k\, =\, 1}^{N}
\sigma_{k} \sigma_{k + 1} \},
\end{equation}
$m,n = 1,...,N$. The definitions (\ref{1.2}),
(\ref{1.9}) and the relation (\ref{1.7}) imply
\begin{equation}
\label{1.10} \sum_{m,n\, =\, 1,...,N,\, \, m\, \neq \, n} <
\sigma_{m} \sigma_{n} >_{N} = \beta^{- 2}
\frac{\partial^{2}}{\partial H^{2}} \left( \ln
\frac{Z_{0}(J,H;T(1,N))}{Z_{0}(0,H;T(1,N))} \right)_{H\, =\, 0}.
\end{equation}
The relations (\ref{1.3}), (\ref{1.7}), (\ref{1.10}) imply
\begin{equation}
\label{1.11} \lim_{N \rightarrow \infty} N^{- 1} \sum_{m,n\, =\,
1,...,N,\, \, m\, \neq \, n}
< \sigma_{m} \sigma_{n} >_{N} \, = \exp \{ 2\beta J\} - 1.
\end{equation}
By making use of the relation (\ref{1.11}) it is possible to express the
spontaneous magnetization (\ref{1.8}) through the two-spin correlation
functions
\begin{eqnarray}
\label{1.12} \lim_{N \rightarrow \infty} \frac{\partial}{\partial x}
\left( \ln \frac{Z_{0}(J,N^{- 1/2}\beta^{- 1} x;T(1,N))}{Z_{0}(0,N^{-
1/2}\beta^{- 1} x;T(1,N))} \right)_{x\, =\, \tanh \beta H} =
\nonumber
\\ (2\tanh \beta H) \lim_{N \rightarrow \infty} N^{- 1} \sum_{m,n\,
=\, 1,...,N,\, \, m\, < \, n}
< \sigma_{m} \sigma_{n} >_{N}.
\end{eqnarray}
The magnetization (\ref{1.5}) expression is cumbersome. For the Ising model
(\ref{1.2}) the right-hand side of the equality of the type (\ref{1.16}) is
equal to zero. For the sufficiently small $\beta J_{1}$ and $\beta J_{2}$
the equality of the type (\ref{1.12}) is proved for the two-dimensional Ising
model in the paper \cite{9}.

In this paper the correlation functions are calculated and the equality of the
type (\ref{1.12}) is proved for the $d$ - dimensional Ising model
($d = 1,2,3$) with the free boundary conditions and for the $d$ - dimensional
${\bf Z}_{2}$ electrodynamics ($d = 2,3$) with the free boundary conditions.
${\bf Z}_{2}$ electrodynamics was introduced in the paper \cite{4}.

\section{Correlation Functions}
\setcounter{equation}{0}

\noindent Let us define Ising model and ${\bf Z}_{2}$
electrodynamics by making use of the algebraic topology notations.
We consider a rectangular lattice formed by the points with integral
Cartesian coordinates $x_{i} = k_{i}$, $M_{i}^{\prime } \leq k_{i}
\leq M_{i}$ , $i = 1,...,d$, $1 \leq d \leq 3$, and the
corresponding edges connecting these vertices. We denote this graph
by $G(M_{1}^{\prime },...,M_{d}^{\prime };M_{1},..,M_{d})$ or simply
$G(M)$. We consider the free boundary conditions. For the periodic
boundary conditions the lattice is wrapped on a torus. The graph
$G(M_{1}^{\prime },...,M_{d}^{\prime};M_{1},..,M_{d})$ cells:
vertices, edges, faces ($d = 2,3$), cubes ($d = 3$) are called the
cells of dimension $0,1,2,3$. They are denoted by
$s_{i}^{0},s_{i}^{1},s_{i}^{2},s_{i}^{3}$.
The cell complex $P(G(M))$ consists of the vertices of the graph
$G(M_{1}^{\prime },...,M_{d}^{\prime};M_{1},..,M_{d})$ and of the
cells of dimension $1,2,3$ whose boundaries contain the cells of
the graph $G(M_{1}^{\prime },...,M_{d}^{\prime};M_{1},..,M_{d})$.
Let ${\bf Z}_{2}^{add} = \{ 0,1 \}$ be the group of modulo $2$
residuals. The modulo $2$ residuals are multiplied by each other
and the group ${\bf Z}_{2}^{add}$ is the field. To every pair of
the cells $s_{i}^{p}$, $s_{j}^{p - 1}$ there corresponds the incidence number
$(s_{i}^{p}:s_{j}^{p - 1}) \in {\bf Z}_{2}^{add}$. If the cell
$s_{j}^{p - 1}$ is included into the boundary of the cell
$s_{i}^{p}$, then the incidence number $(s_{i}^{p}:s_{j}^{p - 1}) =
1 \in {\bf Z}_{2}^{add}$. Otherwise the incidence number
$(s_{i}^{p}:s_{j}^{p - 1}) = 0 \in {\bf Z}_{2}^{add}$. For any pair
of the cells $s_{i}^{p + 1}$, $s_{j}^{p - 1}$ the incidence numbers
satisfy the condition
\begin{equation}
\label{2.1} \sum_{s_{m}^{p}\, \in \, P(G(M))} (s_{i}^{p +
1}:s_{m}^{p})(s_{m}^{p}:s_{j}^{p - 1}) = 0.
\end{equation}
A cochain $c^{p}$ of the cell complex $P(G(M))$ with the
coefficients in the group ${\bf Z}_{2}^{add}$ is a function on the
$p$ - dimensional cells taking values in the group ${\bf
Z}_{2}^{add}$. Usually the oriented cells $\pm s^{p}$ are considered
and the cochains are the antisymmetric functions:
$c^{p}(- s^{p}) = - c^{p}(+ s^{p})$. However, $- 1 = 1 \,
\hbox{mod} \, 2$ and we can neglect the cell orientation for the
coefficients in the group ${\bf Z}_{2}^{add}$:
$c^{p}(- s^{p}) = c^{p}(+ s^{p})$. The cochains form an Abelian
group $C^{p}(P(G(M)),{\bf Z}_{2}^{add})$
\begin{equation}
\label{2.2} (c^{p} + c^{\prime p})(s_{i}^{p}) = c^{p}(s_{i}^{p}) +
c^{\prime p}(s_{i}^{p}).
\end{equation}
The homomorphism
\begin{equation}
\label{2.3}
\partial c^{p}(s_{i}^{p - 1}) = \sum_{s_{j}^{p}\, \in \, P(G(M))}
(s_{j}^{p}:s_{i}^{p - 1}) c^{p}(s_{j}^{p})
\end{equation}
of the group $C^{p}(P(G(M)),{\bf Z}_{2}^{add})$ into the group
$C^{p - 1}(P(G(M)),{\bf Z}_{2}^{add})$ is called the boundary operator.
Let us introduce the bilinear form on the group
$C^{p}(P(G(M)),{\bf Z}_{2}^{add})$:
\begin{equation}
\label{2.5} \langle f^{p},g^{p} \rangle = \sum_{s_{i}^{p} \in
P(G(M))} f^{p}(s_{i}^{p})g^{p}(s_{i}^{p}).
\end{equation}
The homomorphism
\begin{equation}
\label{2.4}
\partial^{\ast} c^{p}(s_{i}^{p + 1}) = \sum_{s_{j}^{p}\, \in \, P(G(M))}
(s_{i}^{p + 1}:s_{j}^{p}) c^{p}(s_{j}^{p})
\end{equation}
of the group $C^{p}(P(G(M)),{\bf Z}_{2}^{add})$ into the group
$C^{p + 1}(P(G(M)),{\bf Z}_{2}^{add})$ is called the
coboundary operator:
\begin{equation}
\label{2.6} \langle f^{p},\partial^{\ast} g^{p - 1} \rangle =
\langle
\partial f^{p},g^{p - 1} \rangle, \, \, \langle f^{p},\partial g^{p
+ 1} \rangle = \langle \partial^{\ast} f^{p},g^{p + 1} \rangle .
\end{equation}
The kernel $Z_{p}(P(G(M)),{\bf Z}_{2}^{add})$ of the homomorphism (\ref{2.3})
is called the group of cycles of the complex $P(G(M))$ with the coefficients
in the group ${\bf Z}_{2}^{add}$. The image $B_{p - 1}(P(G(M)),{\bf Z}_{2}^{add})$
of the homomorphism (\ref{2.3}) is called the group of boundaries
of the complex $P(G(M))$ with the coefficients in the group ${\bf Z}_{2}^{add}$.
The condition (\ref{2.1}) implies $\partial^{\ast} \partial^{\ast}  = 0$
and $\partial \partial = 0$: the group
$B_{p - 1}(P(G(M)),{\bf Z}_{2}^{add})$ is the subgroup of the group
$Z_{p - 1}(P(G(M)),{\bf Z}_{2}^{add})$.

The energy is the function on the cochains $\sigma^{p} \in
C^{p}(P(G(M)),{\bf Z}_{2}^{add})$
\begin{equation}
\label{2.7} \overline{H}_{0}^{\prime }(\partial^{\ast} \sigma^{p} )
= \sum_{s_{i}^{p + 1} \in P(G(M))} h(\partial^{\ast} \sigma^{p}
(s_{i}^{p + 1});s_{i}^{p + 1})
\end{equation}
where an arbitrary function $h(\epsilon ;s_{i}^{p + 1})$ on
${\bf Z}_{2}^{add}$ depends on the cell $s_{i}^{p + 1}$:
\begin{equation}
\label{2.8} h(\epsilon ;s_{i}^{p + 1}) = D(s_{i}^{p + 1}) -
J(s_{i}^{p + 1})(- 1)^{\epsilon},
\end{equation}
$$
D(s_{i}^{p + 1}) = \frac{1}{2} \left( h(1;s_{i}^{p + 1}) +
h(0;s_{i}^{p + 1})\right), \, \, J(s_{i}^{p + 1}) = \frac{1}{2}
\left( h(1;s_{i}^{p + 1}) - h(0;s_{i}^{p + 1})\right).
$$
$\epsilon \rightarrow (- 1)^{\epsilon}$ is the isomorphism of the
additive group ${\bf Z}_{2}^{add} = \{ 0,1\}$ into the multiplicative group
${\bf Z}_{2} = \{ \pm 1 \}$. The substitution of the equality
(\ref{2.8}) into the equality (\ref{2.7}) gives
\begin{equation}
\label{2.10} \overline{H}_{0}^{\prime }(\partial^{\ast} \sigma^{p} )
= D + \overline{H}_{0} (\partial^{\ast} \sigma^{p} ),\, \,
D = \sum_{s_{i}^{p + 1} \in P(G(M))} D(s_{i}^{p + 1}),
\end{equation}
\begin{equation}
\label{2.11} \overline{H}_{0} (\partial^{\ast} \sigma^{p} ) = -
\sum_{s_{i}^{p + 1} \in P(G(M))} J(s_{i}^{p + 1}) (-
1)^{\partial^{\ast} \sigma^{p} (s_{i}^{p + 1})}.
\end{equation}
The number $\partial^{\ast} \sigma^{p}(s_{i}^{p + 1})$ is given by the relation
(\ref{2.4}). Let the interaction energy $J(s_{i}^{1})$ depend on the edge $s_{i}^{1}$
orientation only: $J_{1}$ ($J_{2}$) is $J(s_{i}^{1})$ for the
horizontally (vertically) oriented edges $s_{i}^{1}$. For $d = 2$,
$p = 0$ and for the numbers
$\sigma_{n,m}= (- 1)^{\sigma^{0} (s_{n,m}^{0})} = \pm 1$
the function (\ref{2.11}) is the energy (\ref{1.15}), $H = 0$, for the
two-dimensional Ising model
\begin{equation}
\label{2.111} - J_{1}\sum_{n\, =\, M_{1}^{\prime} - 1}^{M_{1}}
\, \sum_{m\, =\, M_{2}^{\prime}}^{M_{2}} \, \sigma_{n,m}
\sigma_{n + 1,m} - J_{2}\sum_{n\, =\, M_{1}^{\prime}}^{M_{1}}
\, \sum_{m\, =\, M_{2}^{\prime} - 1}^{M_{1}} \, \sigma_{n,m} \sigma_{n,m + 1}.
\end{equation}
There are no vertices $(M_{1}^{\prime} - 1,m)$, $(n,M_{2}^{\prime} - 1)$,
$(M_{1} + 1,m)$, $(n,M_{2} + 1)$ in the cell complex $P(G(M))$.
The values $\sigma_{M_{1}^{\prime} - 1,m}$, $\sigma_{n,M_{2}^{\prime} - 1}$,
$\sigma_{M_{1} + 1,m}$, $\sigma_{n,M_{2} + 1}$ are equal to one. The term
$ - J_{1}\sigma_{M_{1}^{\prime},m}$ in the sum (\ref{2.111}) and other boundary
terms are neglected in the Hamiltonian (\ref{1.15}). For $p = 1$ the function
(\ref{2.11}) is the energy for ${\bf Z}_{2}$ electrodynamics \cite{4}. By
making use of the numbers
$\sigma (s_{i}^{1})= (- 1)^{\sigma^{1} (s_{i}^{1})} = \pm 1$ it is possible
to rewrite the function (\ref{2.11}), $p = 1$ in the form (\ref{2.111}) with
the products of four numbers $\sigma (s_{i}^{1})$. The numbers
${\bf Z}_{2}^{add}$ and the algebraic topology notations allow us to consider
Ising model and ${\bf Z}_{2}$ electrodynamics together.
The Ising model and the ${\bf Z}_{2}$ electrodynamics are the mathematical
models of the ferromagnetic crystals. From the algebraic point of
view these models are similar. The magnetism is connected with the
currents flowing along the closed contours. The expression
(\ref{2.11}) gives the energy of the ${\bf Z}_{2}$ - currents
flowing along the closed contours $\partial s_{i}^{p + 1}$. For the
${\bf Z}_{2}$ electrodynamics ($p = 1$) the closed contour
$\partial s_{i}^{2}$ consists in general of four boundary edges of the face
$s_{i}^{2}$. For the Ising model ($p = 0$) the closed contour $\partial
s_{i}^{1}$ consists in general of two boundary vertices of the edge
$s_{i}^{1}$. It seems that the ${\bf Z}_{2}$ electrodynamics has more
physical sense than the Ising model.

The equality (\ref{2.10}) implies
$$
\sum_{\sigma^{p} \in C^{p}(P(G(M)),{\bf Z}_{2}^{add})}
\exp \{ - \beta \overline{H}_{0}^{\prime }(\partial^{\ast} \sigma^{p} )\}
= Z_{p}(J,0;G(M))\exp \{ - \beta D \},
$$
\begin{equation}
\label{2.13} Z_{p}(J,0;G(M)) = \sum_{\sigma^{p} \in
C^{p}(P(G(M)),{\bf Z}_{2}^{add})} \exp \{ - \beta \overline{H}_{0}
(\partial^{\ast} \sigma^{p} )\}.
\end{equation}
The function (\ref{2.13}) is the partition function of Ising model
($p = 0$) and of ${\bf Z}_{2}$ electrodynamics ($p = 1$) in the absence of
magnetic field.

Let the cochain $\chi^{p} \in C^{p}(P(G(M)),{\bf Z}_{2}^{add})$ take
the value $1 \in {\bf Z}_{2}^{add}$ at the cells
$s_{1}^{p}$,...,$s_{m}^{p}$ and be equal to $0 \in {\bf
Z}_{2}^{add}$ at all other $p$ - dimensional cells of the graph
$G(M)$. The function
$$
\alpha (\chi^{p}; G(M)) =
$$
$$
(Z_{p}(J,0;G(M)) \exp \{ - \beta D \})^{- 1}
\sum_{\sigma^{p} \in C^{p}(P(G(M)),{\bf Z}_{2}^{add})} (-
1)^{\langle \chi^{p} ,\sigma^{p} \rangle }
\exp \{ - \beta \overline{H}_{0}^{\prime }(\partial^{\ast} \sigma^{p} )\} =
$$
\begin{equation}
\label{2.14}
(Z_{p}(J,0;G(M)))^{- 1} \sum_{\sigma^{p} \in C^{p}(P(G(M)),{\bf
Z}_{2}^{add})} (- 1)^{\langle \chi^{p} ,\sigma^{p} \rangle } \exp \{
- \beta \overline{H}_{0} (\partial^{\ast} \sigma^{p} )\},
\end{equation}
$(- 1)^{\langle \chi^{p} ,\sigma^{p} \rangle } = (- 1)^{\sigma^{p} (s_{1}^{p})}
\cdots (- 1)^{\sigma^{p} (s_{m}^{p})}$, is the correlation function at the
cells $s_{1}^{p}$,...,$s_{m}^{p}$ of the lattice $G(M)$. The definitions
(\ref{1.9}) and (\ref{2.14}) are consistent. If the cochain $0$ takes the
value $0 \in {\bf Z}_{2}^{add}$ at any $p$ - dimensional cell of the lattice
$G(M)$, then the correlation function $\alpha (0; G(M)) = 1$. The function
(\ref{2.14}) is the correlation function of Ising model ($p = 0$) and of
${\bf Z}_{2}$ electrodynamics ($p = 1$) in the absence of magnetic field.
For the particular values of the interaction energies the correlation
functions of the three-dimensional ${\bf Z}_{2}$ electrodynamics with free
boundary conditions are calculated in the paper \cite{4}. These correlation
functions are connected with the correlation functions of the
two-dimensional Ising model. Below we calculate the correlation
functions of the Ising model and the ${\bf Z}_{2}$ electrodynamics
for the case when the sign of the interaction energy $J(s_{i}^{p + 1}$)
is independent of the cell $s_{i}^{p + 1}$ and the interaction energy
$J(s_{i}^{p + 1})$ depends on the cell $s_{i}^{p + 1}$. The interaction
energy $J(s_{i}^{p + 1})$ is supposed to be small in contrast with the paper
\cite{7}.

By making use of the harmonic analysis on the group
$C^{p}(P(G(M)),{\bf Z}_{2}^{add})$ and the first relation (\ref{2.6}) it
is possible to prove (\cite{10}, Proposition 3.1)
\begin{equation}
\label{2.15} Z_{p}(J,0;G(M)) = 2^{\# (G;p)} \left( \prod_{s_{i}^{p +
1} \, \in \, P(G(M))} \cosh \beta J(s_{i}^{p + 1})\right)
Z_{r,p}(J,0;G(M)),
\end{equation}
\begin{equation}
\label{2.9} Z_{r,p}(J,0;G(M)) = \sum_{\xi^{p + 1} \, \in \, C^{p +
1}(P(G(M)),{\bf Z}_{2}^{add}),\, \, \partial \xi^{p + 1} \, =\, 0 }
||\xi^{p + 1} ||_{J,G(M)},
\end{equation}
\begin{equation}
\label{2.16} \alpha (\chi^{p}; G(M)) = (Z_{r,p}(J,0;G(M)))^{- 1}
\sum_{\xi^{p + 1} \, \in \, C^{p + 1}(P(G(M)),{\bf Z}_{2}^{add}),\,
\, \partial \xi^{p + 1} \, =\, \chi^{p} } ||\xi^{p + 1} ||_{J,G(M)},
\end{equation}
\begin{equation}
\label{2.160} ||\xi^{p + 1} ||_{J,G(M)} =  \prod_{s_{i}^{p + 1} \,
\in \, P(G(M))} \left( \tanh \beta J(s_{i}^{p + 1})\right)^{\tau ((- 1)^{\xi^{p + 1} (s_{i}^{p + 1})})},
\end{equation}
\begin{equation}
\label{2.12}
\tau ((- 1)^{\epsilon}) = \frac{1 - (- 1)^{\epsilon}}{2} = \left\{ {1,
\hskip 0,5cm \epsilon = 1 \in {\bf Z}_{2}^{add},}
\atop {0, \hskip 0,5cm \epsilon = 0 \in {\bf Z}_{2}^{add}.} \right.
\end{equation}
The constant $\# (G(M);p)$ is the total number of the $p$ - dimensional
cells of the cell complex $P(G(M))$. The correlation function (\ref{2.16})
is equal to zero for $\chi^{p} \notin B_{p}(P(G(M)),{\bf Z}_{2}^{add})$.

For the cochain
$\chi^{0} \in B_{0}(P(G(M_{1}^{\prime };M_{1}))),{\bf Z}_{2}^{add})$
the equation $\partial \xi^{1} = \chi^{0}$ has the unique solution.
The group of cycles
$Z_{1}(P(G(M_{1}^{\prime };M_{1})),{\bf Z}_{2}^{add})$ consists of
the cochain $0$. The equalities (\ref{2.9}), (\ref{2.16}) imply
\begin{equation}
\label{2.1611} Z_{r,0}(J,0;G(M_{1}^{\prime };M_{1})) = 1,
\end{equation}
\begin{equation}
\label{2.1612} \alpha (\chi^{0}; G(M_{1}^{\prime };M_{1})) =
||\xi^{1} ||_{J,G(M_{1}^{\prime };M_{1})}.
\end{equation}
For the cochain
$\chi^{0} \notin B_{0}(P(G(M_{1}^{\prime };M_{1})),{\bf Z}_{2}^{add})$
the equation $\partial \xi^{1} = \chi^{0}$ has no solutions. The
equality (\ref{2.16}) implies
\begin{equation}
\label{2.1613} \alpha (\chi^{0}; G(M_{1}^{\prime };M_{1})) = 0,\, \,
\chi^{0} \notin B_{0}(P(G(M_{1}^{\prime };M_{1})),{\bf Z}_{2}^{add}).
\end{equation}
For the cochain $\chi^{1} \in$
$B_{1}(P(G(M_{1}^{\prime },M_{2}^{\prime };M_{1},M_{2})),{\bf
Z}_{2}^{add})$. The equation $\partial \xi^{2} = \chi^{1}$ has
unique solution. The group of cycles $Z_{2}(P(G(M_{1}^{\prime },
M_{2}^{\prime}; M_{1},M_{2})),{\bf Z}_{2}^{add})$ consists of the
cochain $0$. The equalities (\ref{2.9}), (\ref{2.16}) imply
\begin{equation}
\label{2.1621} Z_{r,1}(J,0;G(M_{1}^{\prime },M_{2}^{\prime
};M_{1},M_{2})) = 1,
\end{equation}
\begin{equation}
\label{2.1622} \alpha (\chi^{1}; G(M_{1}^{\prime },M_{2}^{\prime
};M_{1},M_{2})) = ||\xi^{2} ||_{J,G(M_{1}^{\prime },M_{2}^{\prime
};M_{1},M_{2})}.
\end{equation}
For the cochain $\chi^{1} \notin$
$B_{1}(P(G(M_{1}^{\prime },M_{2}^{\prime };M_{1},M_{2})),{\bf
Z}_{2}^{add})$ the equation $\partial \xi^{2} = \chi^{1}$ has no
solutions. The equality (\ref{2.16}) implies
\begin{equation}
\label{2.1623}
\alpha (\chi^{1}; G(M_{1}^{\prime
},M_{2}^{\prime };M_{1},M_{2})) = 0,\, \, \chi^{1} \notin
B_{1}(P(G(M_{1}^{\prime },M_{2}^{\prime };M_{1},M_{2})),{\bf
Z}_{2}^{add}).
\end{equation}

The partition function (\ref{2.15}) for $p = 0$, $d = 2$ was
"obtained" by Kac and Ward \cite{11}:

\noindent "The partition function of the two-dimensional square net
Ising model can be easily put in the form \cite{12}
\begin{equation}
\label{2.163} (\cosh \beta J_{2})^{h}(\cosh \beta J_{1})^{v} \sum
g(l,k)x^{l}y^{k},
\end{equation}
where
$$
x = \tanh \beta J_{2},\, \, y = \tanh \beta J_{1},
$$
$h$ the total number of horizontal links, $v$ the total number of
vertical links, and $g(l,k)$ the number of "closed polygons" with
$l$ horizontal and $k$ vertical links."

Let us compare the expression (\ref{2.15}) with the expression
(\ref{2.163}):
$$
(\cosh \beta J_{2})^{h}(\cosh \beta J_{1})^{v} \sum g(l,k)(\tanh
\beta J_{2})^{l}(\tanh \beta J_{1})^{k} =
$$
\begin{equation}
\label{2.164}
\left( \prod_{s_{i}^{p + 1} \, \in \, P(G(M))} \cosh \beta
J(s_{i}^{p + 1})\right) \sum_{\xi^{1} \, \in \, C^{1}(P(G(M)),{\bf
Z}_{2}^{add}),\, \,
\partial \xi^{1} \, =\, 0 } ||\xi^{1} ||_{J,G(M)}.
\end{equation}
The interaction energy $J(s_{i}^{1})$ depends on the orientation of
the edge $s_{i}^{1}$ only. The normalization constant
$2^{\# (G(M);p)} = (\# \{ \epsilon \in {\bf Z}_{2}^{add} \} )^{\# (G(M);p)}$
for the harmonic analysis on the group $C^{p}(P(G(M)),{\bf Z}_{2}^{add})$
is  missed in the expression (\ref{2.164}).

There is no any expression for the partition function of Ising model
in the paper \cite{12}. Van der Waerden believed that the sum with the
"long order" \cite{12}
\begin{equation}
\label{2.165}
\sum g(l,k)z^{l + k}, \, \, z = \exp \{ - \beta J \}
\neq \tanh \beta J
\end{equation}
is important to study the crystals. $g(l,k)$ is the number of closed
polygons with $l$ horizontal and $k$ vertical links. It seems that the
definitions of the number $g(l,k)$ in the papers \cite{11} and \cite{12}
are different. Van der Waerden did not use the modulo 2 residuals. Kac,
Ward \cite{11} and van der Waerden \cite{12} did not consider the
correlation functions and avoided to use the algebraic topology
notations. The relation (\ref{2.16}) needs the algebraic topology
notations.

Let the edge $\{ (k_{1},k_{2}),(l_{1},l_{2})\}$ have the end
vertices $(k_{1},k_{2})$, $(l_{1},l_{2}) \in$ $G(M_{1}^{\prime
},M_{2}^{\prime };M_{1},M_{2})$. The oriented edge
$((k_{1},k_{2}),(l_{1},l_{2}))$ has the initial vertex
$(k_{1},k_{2})$ and the final vertex $(l_{1},l_{2})$. Kac and Ward
\cite{11}: "In the main body of the paper we shall explain in detail
the method of computing which yields the partition function up to
negligible terms due to boundary effects. Several combinatorial
points will be dealt with a heuristic manner only. We do not go into
the details of rigor because our main aim is not so much an
alternative derivation of the Onsager-Kaufman formula but a
demonstration that a combinatorial approach is indeed possible."
Kac and Ward \cite{11} discussed the following formula for the
partition function (\ref{2.9})
\begin{equation}
\label{2.166} Z_{r,0}(J,0;G(M_{1}^{\prime },M_{2}^{\prime };M_{1},M_{2})) = \det (I + T(J))
\end{equation}
where $I$ is the identity matrix on the set of the oriented edges
$((k_{1},k_{2}),(l_{1},l_{2}))$ and the interaction matrix
$$
T(J)_{(((k_{1},k_{2}),(l_{1},l_{2})),((k_{1}^{\prime},k_{2}^{\prime}),
(l_{1}^{\prime},l_{2}^{\prime})))} = 0,\, \, (l_{1},l_{2}) \neq
(k_{1}^{\prime},k_{2}^{\prime}),
$$
$$
T(J)_{(((k_{1},k_{2}),(l_{1},l_{2})),((l_{1},l_{2}),(k_{1}^{\prime},k_{2}^{\prime})))}
= 0,\, \, (k_{1},k_{2}) = (k_{1}^{\prime},k_{2}^{\prime}),
$$
$$
T(J)_{(((k_{1},k_{2}),(l_{1},l_{2})),((l_{1},l_{2}),(k_{1}^{\prime},k_{2}^{\prime})))}
= \tanh (\beta J(\{ (k_{1},k_{2}),(l_{1},l_{2})\})) \times
$$
\begin{equation}
\label{2.17}
\exp \Bigl\{ \frac{i}{2} \, \, [ (l_{1} - k_{1},l_{2} -
k_{2}),(k_{1}^{\prime} - l_{1},k_{2}^{\prime} - l_{2})] \Bigr\}, \, \,
(k_{1},k_{2}) \neq (k_{1}^{\prime},k_{2}^{\prime}).
\end{equation}
For a vertical edge $J(\{ (k_{1},k_{2}),(k_{1},k_{2} + 1)\}) =
J_{1}$ and for a horizontal edge $J(\{ (k_{1},k_{2}),(k_{1} +
1,k_{2})\}) = J_{2}$. The number
$[(l_{1},l_{2}),(l_{1}^{\prime},l_{2}^{\prime})]$ is the minimal radian
measure of the angle between the direction of the vector
$(l_{1},l_{2})$ and the direction of the vector
$(l_{1}^{\prime},l_{2}^{\prime})$. For an arbitrary finite connected
graph $G$ on the lattice ${\bf Z}^{\times 2}$ the following formula
\begin{equation}
\label{2.18} Z_{r,0}^{2}(J,0;G) = \det (I - T(J))
\end{equation}
is proved in the paper \cite{15}. By making use of the formulae
(\ref{2.16}), (\ref{2.18}) the correlation functions of the
two-dimensional Ising model with the free boundary conditions are
obtained in the paper \cite{10}. The formula (\ref{2.18}) implies
the alternative derivation \cite{10} of the Onsager-Kaufman formula.
For the periodic boundary conditions McCoy and Wu \cite{8}
represented the partition function (\ref{2.9}) as the linear
combination of Pfaffians. The counterexample for the McCoy-Wu formula
\cite{8} was constructed in the paper \cite{16}. The Euler
characteristic of the orientable two-dimensional sphere $S^{2}$ is
equal to $2$. It implies the simple proof of the formula (\ref{2.18})
in the paper \cite{16}. The definition (\ref{2.17}) uses the plane
lattice crucially. The formula (\ref{2.18}) for the three-dimensional
Ising model is not clear.

The spontaneous magnetization (\ref{1.12}) depends on the correlation
functions only. By making use of the formula (\ref{2.16}) we shall obtain
the correlation functions of the $d$ - dimensional Ising model ($d = 2,3$)
and of the three-dimensional ${\bf Z}_{2}$ electrodynamics with the
free boundary conditions without calculation of the partition functions
(\ref{2.9}). In order to calculate the correlation functions (\ref{2.16})
we need the notion of the connected cochain.

The set of the cells $s_{i}^{p + 1}$ on which the cochain
$\zeta^{p + 1} \in C^{p + 1}(P(G(M)),{\bf Z}_{2}^{add})$ takes the value $1
\in {\bf Z}_{2}^{add}$ is called the support of the cochain $\zeta^{p + 1}$.
The nonzero cochain $\zeta^{p + 1} \in $ $C^{p + 1}(P(G(M)),{\bf Z}_{2}^{add})$
is called connected if for any two cells $s_{i}^{p + 1}$,
$s_{j}^{p + 1}$  from the support of the cochain $\zeta^{p + 1}$ there exists a
connecting sequence of the cells $s_{i}^{p + 1} = s_{1}^{p + 1}$, $s_{2}^{p + 1}$,
...,$s_{l - 1}^{p + 1}$, $s_{l}^{p + 1} = s_{j}^{p + 1}$ from the
support of the cochain $\zeta^{p + 1}$ with the common boundary cells:
\begin{equation}
\label{2.19}
s_{k}^{p}: \, \, (s_{k}^{p + 1}:s_{k}^{p}) (s_{k + 1}^{p + 1}:s_{k}^{p})
= 1 \in {\bf Z}_{2}^{add}, \, \, k = 1,...,l - 1.
\end{equation}
Any nonzero cochain
$\zeta^{p + 1} \in C^{p + 1}(P(G(M)),{\bf Z}_{2}^{add})$ can be uniquely
represented as the sum of the connected nonzero cochains $\zeta^{p +
1}_{m} \in C^{p + 1}(P(G(M)),{\bf Z}_{2}^{add})$, $m = 1,...,k$. For $m \neq n$
the cells from the supports of the cochains $\zeta^{p + 1}_{m}$,
$\zeta^{p + 1}_{n}$ have no the common boundary cells:
\begin{equation}
\label{2.20}
(s_{i}^{p + 1}:s_{k}^{p}) (s_{j}^{p + 1}:s_{k}^{p}) = 0 \in {\bf Z}_{2}^{add},
\, \, \zeta^{p +
1}_{m} (s_{i}^{p + 1}) = 1,\, \, \zeta^{p +
1}_{n} (s_{j}^{p + 1}) = 1, \, \, m \neq n,
\end{equation}
$m,n = 1,...,k$. For the nonzero cochain $\chi^{p} \in B_{p}(P(G(M)),{\bf Z}_{2}^{add})$
any solution of the equation $\partial \xi^{p + 1} (M) = \chi^{p}$ can be
uniquely represented as
\begin{equation}
\label{2.23}
\xi^{p + 1} = \sum_{i_{1}\, =\, 1}^{k_{1}} \lambda_{1i_{1}}^{p + 1} (M) + \xi_{1}^{p + 1}.
\end{equation}
The connected cochains $\lambda_{1i_{1}}^{p + 1} (M)
\in C^{p + 1}(P(G(M)),{\bf Z}_{2}^{add})$, $i_{1} = 1,...,k_{1}$, satisfy the equations
$$
\partial \left( \sum_{i_{1}\, =\, 1}^{k_{1}} \lambda_{1i_{1}}^{p + 1} (M) \right) = \chi^{p}; \, \,
\partial \lambda_{1i_{1}}^{p + 1} (M) \neq 0;
$$
\begin{equation}
\label{2.25}
(s_{i}^{p + 1}:s_{l}^{p}) (s_{j}^{p + 1}:s_{l}^{p}) = 0, \, \,
\lambda_{1i_{1}}^{p + 1} (M) (s_{i}^{p + 1}) = 1,\, \,
\lambda_{1j_{1}}^{p + 1} (M) (s_{j}^{p + 1}) = 1,\, \, i_{1} \neq j_{1},
\end{equation}
$i_{1},j_{1} = 1,...,k_{1}$. The cochain $\xi_{1}^{p + 1} \in C^{p + 1}(P(G(M)),{\bf Z}_{2}^{add})$
satisfies  the equations
$$
\partial \xi_{1}^{p + 1} = 0;
$$
\begin{equation}
\label{2.26}
(s_{i}^{p + 1}:s_{l}^{p}) (s_{j}^{p + 1}:s_{l}^{p}) = 0, \, \,
\xi_{1}^{p + 1} (s_{i}^{p + 1}) = 1,\, \,
\lambda_{1i_{1}}^{p + 1} (M) (s_{j}^{p + 1}) = 1,\, \,
i_{1} = 1,...,k_{1}.
\end{equation}
The cochain $\chi^{p}$ does not determine the integer $k_{1}$ in the relations
(\ref{2.23}) - (\ref{2.26}). The integer $k_{1}$ does not exceed the total number
of cells in the support of the cochain $\chi^{p}$. The relations (\ref{2.160}),
(\ref{2.25}), (\ref{2.26}) imply
\begin{equation}
\label{2.27} ||\sum_{i_{1}\, =\, 1}^{k_{1}} \lambda_{1i_{1}}^{p + 1} (M) +
\xi_{1}^{p + 1} ||_{J,G(M)} = \left( \prod_{i_{1}\, =\, 1}^{k_{1}}
||\lambda_{1i_{1}}^{p + 1} (M)||_{J,G(M)} \right) ||\xi_{1}^{p + 1}
||_{J,G(M)}.
\end{equation}
In view of the equality (\ref{2.23}) the relation (\ref{2.16}) may be rewritten
for the nonzero cochain $\chi^{p} \in B_{p + 1}(P(G(M)),{\bf Z}_{2}^{add})$  as
$$
\alpha (\chi^{p}; G(M)) =  \sum_{1\, \leq \, k_{1}} \, \, \sum_{\lambda_{1i_{1}}^{p +
1} (M):\, \, (\ref{2.25})} \left( \prod_{i_{1}\, =\, 1}^{k_{1}}
||\lambda_{1i_{1}}^{p + 1} (M)||_{J,G(M)} \right)
$$
\begin{equation}
\label{2.28}
\times \left( 1 + \alpha \left( \chi^{p}, \sum_{i_{1}\, =\, 1}^{k_{1}} \lambda_{1i_{1}}^{p
+ 1} (M); G(M)\right) \right)^{- 1},
\end{equation}
$$
\left( 1 + \alpha \left( \chi^{p}, \sum_{i_{1}\, =\, 1}^{k_{1}}
\lambda_{1i_{1}}^{p + 1} (M); G(M)\right) \right)^{- 1} =
$$
\begin{equation}
\label{2.281} (Z_{r,p}
(J,0;G(M)))^{- 1} \sum_{\xi_{1}^{p + 1}:\, \, (\ref{2.26})}
||\xi_{1}^{p + 1} ||_{J,G(M)}.
\end{equation}
The cochain $\xi^{p + 1} \in  Z_{p + 1}(P(G(M)),{\bf Z}_{2}^{add})$
may have the form similar to the sum (\ref{2.23})
\begin{equation}
\label{2.282}
\xi^{p + 1} = \sum_{i_{2}\, =\, 1}^{k_{2}} \lambda_{2i_{2}}^{p + 1} (M) + \xi_{2}^{p + 1}.
\end{equation}
The connected cochains $\lambda_{2i_{2}}^{p + 1} (M) \in  $
$C^{p + 1}(P(G(M)),{\bf Z}_{2}^{add})$, $i_{2} = 1,...,k_{2}$, satisfy the
equations
$$
\partial \lambda_{2i_{2}}^{p + 1} (M) = 0;
$$
$$
\forall \, i_{2}\, \, \exists \, i_{1},i,j,l:\, \, (s_{i}^{p + 1}:s_{l}^{p})
(s_{j}^{p + 1}:s_{l}^{p}) = 1,\, \, \lambda_{2i_{2}}^{p + 1} (M) (s_{i}^{p + 1}) = 1, \, \,
\lambda_{1i_{1}}^{p + 1} (M) (s_{j}^{p + 1}) = 1;
$$
\begin{equation}
\label{2.29} (s_{i}^{p + 1}:s_{l}^{p}) (s_{j}^{p + 1}:s_{l}^{p}) = 0,
\, \, \lambda_{2i_{2}}^{p + 1} (M) (s_{i}^{p + 1}) = 1,\, \,
\lambda_{2j_{2}}^{p + 1} (M) (s_{j}^{p + 1}) = 1,\, \, i_{2} \neq j_{2},
\end{equation}
$i_{2},j_{2} = 1,...,k_{2}$. The cochain
$\xi_{2}^{p + 1} \in$ $C^{p + 1}(P(G(M)),{\bf Z}_{2}^{add})$
satisfies the equations
$$
\partial \xi_{2}^{p + 1} = 0;\, \, (s_{i}^{p + 1}:s_{l}^{p}) (s_{j}^{p + 1}:s_{l}^{p}) = 0,
$$
\begin{equation}
\label{2.30}
\xi_{2}^{p + 1} (s_{i}^{p + 1}) = 1,\, \,
\lambda_{ni_{n}}^{p + 1} (M) (s_{j}^{p + 1}) = 1,\, \,
i_{n} = 1,...,k_{n},\, \, n = 1,2.
\end{equation}
If the cochains $\lambda_{2i_{2}}^{p + 1} (M) = 0$, then the equations (\ref{2.30})
coincide with the equations (\ref{2.26}) and the cochain  (\ref{2.282}) coincides with
the cochain $\xi_{1}^{p + 1}$  satisfying the equations (\ref{2.26}).. The relations
(\ref{2.29}), (\ref{2.30}) imply for the cochain (\ref{2.282}) the relation similar to
the relation (\ref{2.27}). The relations (\ref{2.9}), (\ref{2.26}), (\ref{2.281}) -
(\ref{2.30}) imply
$$
Z_{r,p} (J,0;G(M)) = \sum_{\xi_{1}^{p + 1}:
(\ref{2.26})} ||\xi_{1}^{p + 1} ||_{J,G(M)} +
$$
\begin{equation}
\label{2.32}
\sum_{1\, \leq \, k_{2}} \, \,
\sum_{\lambda_{2i_{2}}^{p + 1} (M): (\ref{2.29})}
\left( \prod_{i_{2}\, =\, 1}^{k_{2}} ||\lambda_{2i_{2}}^{p + 1} (M)||_{J,G(M)}
\right) \sum_{\xi_{2}^{p + 1}:\, \, (\ref{2.30})} ||\xi_{2}^{p + 1}
||_{J,G(M)},
\end{equation}
$$
\alpha \left( \chi^{p}, \sum_{i_{1}\, =\, 1}^{k_{1}}
\lambda_{1i_{1}}^{p + 1} (M); G(M)\right) = \sum_{1\, \leq \, k_{2}}
\, \, \sum_{\lambda_{2i_{2}}^{p + 1} (M):\, \, (\ref{2.29})}
\left( \prod_{i_{2}\, =\, 1}^{k_{2}}
||\lambda_{2i_{2}}^{p + 1} (M)||_{J,G(M)} \right)
$$
\begin{equation}
\label{2.33}
\times \left( 1 + \alpha \left( \chi^{p}, \sum_{i_{1}\, =\, 1}^{k_{1}} \lambda_{1i_{1}}^{p
+ 1} (M), \sum_{i_{2}\, =\, 1}^{k_{2}} \lambda_{2i_{2}}^{p + 1} (M); G(M)\right)
\right)^{- 1},
\end{equation}
$$
\left( 1 + \alpha \left( \chi^{p}, \sum_{i_{1}\, =\, 1}^{k_{1}}
\lambda_{1i_{1}}^{p + 1} (M), \sum_{i_{2}\, =\, 1}^{k_{2}} \lambda_{2i_{2}}^{p + 1}
(M);G(M)\right) \right)^{- 1} =
$$
\begin{equation}
\label{2.331}
\left( \sum_{\xi_{1}^{p + 1}: \, \, (\ref{2.26})} ||\xi_{1}^{p + 1}
||_{J,G(M)} \right)^{- 1} \sum_{\xi_{2}^{p + 1}: \, \, (\ref{2.30})}
||\xi_{2}^{p + 1} ||_{J,G(M)}.
\end{equation}
We continue this process to construct the sequence of the connected
cochains  $\lambda_{ni_{n}}^{p + 1} (M)$, $i_{n} = 1,...,k_{n}$, $n = 1,2,...$,
from the group $C^{p + 1}(P(G(M)),{\bf Z}_{2}^{add})$ satisfying the
equations (\ref{2.25}) for $n = 1$, (\ref{2.29}) for $n = 2$ and the equations
$$
(s_{i}^{p + 1}:s_{l}^{p}) (s_{j}^{p + 1}:s_{l}^{p}) = 0,
\, \, \lambda_{ni_{n}}^{p + 1} (M) (s_{i}^{p + 1}) = 1, \, \,
\lambda_{mi_{m}}^{p + 1} (M) (s_{j}^{p + 1}) = 1,
$$
$$
i_{n} = 1,...,k_{n},\, \, i_{m} = 1,...,k_{m},\, \,
m = 1,...,n - 2 \geq 1;
$$
$$
\forall i_{n}\, \, \exists i_{n - 1},i,j,l:\, \, (s_{i}^{p + 1}:s_{l}^{p})
(s_{j}^{p + 1}:s_{l}^{p}) = 1,
$$
$$
\lambda_{ni_{n}}^{p + 1} (M) (s_{i}^{p + 1}) = 1, \, \,
\lambda_{(n - 1)i_{n - 1}}^{p + 1} (M) (s_{j}^{p + 1}) = 1, \, \, n \geq 2;
$$
$$
\partial \lambda_{ni_{n}}^{p + 1} (M) = 0;\, \, (s_{i}^{p + 1}:s_{l}^{p})
(s_{j}^{p + 1}:s_{l}^{p}) = 0,
$$
\begin{equation}
\label{2.34}
\lambda_{ni_{n}}^{p + 1} (M) (s_{i}^{p + 1}) = 1, \, \,
\lambda_{nj_{n}}^{p + 1} (M) (s_{j}^{p + 1}) = 1, \, \, i_{n} \neq j_{n},\, \,
i_{n},j_{n} = 1,...,k_{n},\, \, n \geq 2.
\end{equation}
We construct also the sequence of the cochains $\xi_{n}^{p + 1}$,
$n = 1,2,...$, satisfying the equations (\ref{2.26}) for $n = 1$,
(\ref{2.30}) for $n = 2$ and the equations
$$
\partial \xi_{n}^{p + 1} = 0;\, \, (s_{i}^{p + 1}:s_{l}^{p}) (s_{j}^{p + 1}:s_{l}^{p}) = 0,
$$
\begin{equation}
\label{2.35}
\xi_{n}^{p + 1} (M) (s_{i}^{p + 1}) = 1, \, \,
\lambda_{mi_{m}}^{p + 1} (M) (s_{j}^{p + 1}) = 1,
\, \, i_{m} = 1,...,k_{m},\, \, m = 1,...,n \geq 1.
\end{equation}
Similarly to the relations (\ref{2.33}), (\ref{2.331}) we have the
anti-recurrent relations for $n \geq 2$
$$
\alpha \left( \chi^{p}, \sum_{i_{1}\, =\, 1}^{k_{1}}
\lambda_{1i_{1}}^{p + 1} (M),..., \sum_{i_{n}\, =\, 1}^{k_{n}} \lambda_{ni_{n}}^{p +
1} (M); G(M)\right) =
$$
$$
\sum_{1\, \leq \, k_{n + 1}} \, \, \sum_{{\lambda_{(n + 1)i_{n +
1}}^{p + 1} (M):} \atop {(\ref{2.34}),\, \, n \, \rightarrow \, n + 1}}
\left( \prod_{i_{n + 1}\, =\, 1}^{k_{n + 1}} ||
\lambda_{(n + 1)i_{n + 1}}^{p + 1} (M)||_{P(G(M))} \right)
$$
\begin{equation}
\label{2.36}
\times  \left( 1 + \alpha
\left( \chi^{p}, \sum_{i_{1}\, =\, 1}^{k_{1}} \lambda_{1i_{1}}^{p + 1} (M),...,
\sum_{i_{n + 1}\, =\, 1}^{k_{n + 1}} \lambda_{(n + 1)i_{n + 1} }^{p + 1} (M);
G(M)\right) \right)^{- 1},
\end{equation}
$$
\left( 1 + \alpha \left( \chi^{p}, \sum_{i_{1}\, =\, 1}^{k_{1}}
\lambda_{1i_{1}}^{p + 1} (M),..., \sum_{i_{n + 1}\, =\, 1}^{k_{n + 1}} \lambda_{(n +
1)i_{n + 1} }^{p + 1} (M); G(M)\right) \right)^{- 1} =
$$
\begin{equation}
\label{2.361}
\left( \sum_{\xi_{n}^{p + 1}: \, \, (\ref{2.35})} ||\xi_{n}^{p + 1}
||_{J,G(M)} \right)^{- 1} \sum_{\xi_{n + 1}^{p + 1}: \, \,
(\ref{2.35}),\, \, n \, \rightarrow \, n + 1} ||\xi_{n + 1}^{p + 1}
||_{J,G(M)}.
\end{equation}
For the finite graph $G(M)$ the group $C^{p + 1}(P(G(M)),{\bf Z}_{2}^{add})$
contains the finite number of the cochains. Let $N(\lambda, G(M)) + 1$ be the
maximal number of the cochain in the cochain sequence satisfying the equations
(\ref{2.25}), (\ref{2.29}), (\ref{2.34}). Let $\lambda_{N(\lambda, G(M)) + 2}^{p + 1} (M)$
be the connected cochain satisfying the first and the third equations
(\ref{2.34}). If the second equality (\ref{2.34}) for
$n = N(\lambda, G(M)) + 2$ holds, then we have constructed the
sequence of the cochains $\lambda_{1i_{1}}^{p + 1} (M),...,$
$\lambda_{(n + 1)i_{n + 1}}^{p + 1} (M)$, $\lambda_{n + 2}^{p + 1} (M)$,
$n = N(\lambda, G(M))$, satisfying the equations (\ref{2.25}), (\ref{2.29}),
(\ref{2.34}). This sequence of the cochains does not exist. Therefore
the second equality (\ref{2.34}) for $n = N(\lambda, G(M)) + 2$ is not
valid and the group of the cochains (\ref{2.35}) for
$n = N(\lambda, G(M))$ coincides with the group of the cochains (\ref{2.35}) for
$n = N(\lambda, G(M)) + 1$. Now the relation (\ref{2.361}) implies
\begin{equation}
\label{2.37} \alpha \left( \chi^{p}, \sum_{i_{1}\, =\, 1}^{k_{1}}
\lambda_{1i_{1}}^{p + 1} (M),...,
\sum_{i_{n + 1}\, =\, 1}^{k_{n + 1}}
\lambda_{(n + 1)i_{n + 1}}^{p + 1}
(M); G(M)\right) = 0,\, \, n = N(\lambda, G(M)).
\end{equation}
The anti-recurrent relations (\ref{2.28}), (\ref{2.33}), (\ref{2.36}),
(\ref{2.37}) define the correlation functions for the finite graph $G(M)$.

The length of $\xi^{p + 1} \in C^{p + 1}(P(G(M)),{\bf Z}_{2}^{add})$
is the number of the cells in the support
\begin{equation}
\label{2.38} |\xi^{p + 1}|_{G(M)} = \sum_{s_{i}^{p + 1} \, \in \,
P(G(M))} \tau ((- 1)^{\xi^{p + 1} (s_{i}^{p + 1})}).
\end{equation}
$\tau ((- 1)^{\epsilon})$ is given by the definition (\ref{2.12}). The
homology group triviality  implies the coincidence of the groups
$Z_{p + 1}(P(G(M)),{\bf Z}_{2}^{add})$ and
$B_{p + 1}(P(G(M)),{\bf Z}_{2}^{add})$ for the graph $G(M) =$

\noindent $G(M_{1}^{\prime},...,M_{d}^{\prime};M_{1},..,M_{d})$. Let us compute
the parity of the number (\ref{2.38}) for the cochain $\partial \xi^{p + 2} \in
B_{p + 1}(P(G(M)),{\bf Z}_{2}^{add})$. Any cell $s_{i}^{p + 2}$ from
the support of the cochain $\xi^{p + 2}$ has $2(p + 2)$ boundary
cells $s_{j}^{p + 1}$. Let the cell $s_{j}^{p + 1}$ belong to the
boundaries of $2m + 1$ cells $s_{i}^{p + 2}$ from the support of the
cochain $\xi^{p + 2}$. In order to get the number
$2(p + 2)|\xi^{p + 2}|_{P(G(M))}$ we count the cell $s_{j}^{p + 1}$ exactly
$2m + 1$ times. The cell $s_{j}^{p + 1}$ should be included into the support
of the cochain $\partial \xi^{p + 2}$. Let the cell $s_{j}^{p + 1}$ belong
to the boundaries of $2m$ cells $s_{i}^{p + 2}$ from the support of the
cochain $\xi^{p + 2}$. In order to get the number $2(p + 2)|\xi^{p + 2}|_{P(G(M))}$
we count the cell $s_{j}^{p + 1}$ exactly $2m$ times. The cell $s_{j}^{p + 1}$
should be excluded from the support of the cochain $\partial \xi^{p + 2}$.
The parities of the numbers $|\partial \xi^{p + 2}|_{G(M)}$ and
$2(p + 2)|\xi^{p + 2}|_{P(G(M))}$ coincide
\begin{equation}
\label{2.43} |\partial \xi^{p + 2}|_{G(M)} = 0\, \, \hbox{mod} \, 2.
\end{equation}
Let the sign of the interaction energy $J(s_{i}^{p + 1})$ be
independent of the cell $s_{i}^{p + 1}$. The equalities
(\ref{2.160}), (\ref{2.43}) for a cochain $\xi^{p + 1} \in Z_{p + 1}(P(G(M)),{\bf
Z}_{2}^{add}) = B_{p + 1}(P(G(M)),{\bf Z}_{2}^{add})$ imply
\begin{equation}
\label{2.44} ||\xi^{p + 1} ||_{J,G(M)} = \prod_{s_{i}^{p + 1} \,
\in \, P(G(M))} \left( \tanh \beta J(s_{i}^{p + 1})\right)^{\tau (
(- 1)^{\xi^{p + 1} (s_{i}^{p + 1})})} \geq 0.
\end{equation}
$\tau ((- 1)^{\epsilon})$ is given by the definition (\ref{2.12}). The
definitions (\ref{2.33}), (\ref{2.36}), (\ref{2.37}) and the inequality
(\ref{2.44}) imply
\begin{equation}
\label{2.45} \alpha \left( \chi^{p}, \sum_{i_{1}\, =\, 1}^{k_{1}}
\lambda_{1i_{1}}^{p + 1} (M),..., \sum_{i_{n}\, =\, 1}^{k_{n}} \lambda_{ni_{n}}^{p +
1} (M); G(M)\right) \geq 0, \, \, n = 1,...,N(\lambda, G(M)) + 1.
\end{equation}

Let us estimate the number of the connected cochains $\lambda^{p + 1}$ with
the value $\lambda^{p + 1} (s_{j}^{p + 1}) = 1 \in {\bf Z}_{2}^{add}$ at the
fixed cell $s_{j}^{p + 1}$. Let $s_{l}^{p}$ be a boundary cell of the cell
$s_{j}^{p + 1}$. In order to construct a new $(p + 1)$ - dimensional cell of
the graph $G(M_{1}^{\prime },...,M_{d}^{\prime };M_{1},..,M_{d})$ with the
boundary cell $s_{l}^{p}$ we need to choose a vertex of the cell
$s_{l}^{p}$ and one of $2(d - p) - 1$ edges orthogonal to the cell
$s_{l}^{p}$. One edge orthogonal to the cell $s_{l}^{p}$ corresponds
with the fixed cell $s_{j}^{p + 1}$.  Any number $1,..., 2(d - p) - 1$
of the new $(p + 1)$ - dimensional cells with the boundary cell
$s_{l}^{p}$ may belong to the support of the cohain $\lambda^{p + 1}$.
Due to the Newton binomial formula the possible number of these sets
of the cells from the support of the cohain $\lambda^{p + 1}$ is equal to
\begin{equation}
\label{2.39} \sum_{k\, =\, 0}^{2(d - p) - 1}
\frac{(2(d - p) - 1)!}{k!(2(d - p) - 1 - k)!}
- 1 = 2^{2(d - p) - 1} - 1.
\end{equation}
The number (\ref{2.39}) implies the estimation
\begin{equation}
\label{2.40} \# \{ \hbox{connected} \, \, \lambda^{p + 1}: \lambda^{p + 1} (s_{j}^{p + 1}) = 1
\} < 2^{(2(d - p) - 1) (|\lambda^{p + 1}|_{G(M)} - 1)}.
\end{equation}
If the sign of the interaction energy $J(s_{i}^{p + 1}$) is
independent of the cell $s_{i}^{p + 1}$ and the interaction energy
$J(s_{i}^{p + 1})$ satisfies the inequality
\begin{equation}
\label{2.46} |\tanh \beta J(s_{i}^{p + 1})| < 2^{2(p - d) + 1},
\end{equation}
then the inequalities (\ref{2.45}), (\ref{2.40}) imply that the
sums (\ref{2.28}), (\ref{2.33}) and (\ref{2.36}) are
bounded by the constants independent of the graph $G(M)$.

Let us prove that the sequence of the correlation functions $\alpha
(\chi^{p}; G(M))$ is the convergent Cauchy sequence when $G(M)
\rightarrow {\bf Z}^{\times d}$. Let the graph $G(N_{1}^{\prime
},...,N_{d}^{\prime };N_{1},..,N_{d})$ be the subset of the graph
$G(M_{1}^{\prime },...,M_{d}^{\prime };M_{1},..,M_{d})$:
$M_{i}^{\prime } < N_{i}^{\prime}$, $N_{i} < M_{i}$, $i = 1,...,d$.
The connected cochain $\lambda_{ni_{n}}^{p + 1} (M) \in  C^{p +
1}(P(G(M)),{\bf Z}_{2}^{add})$ coincides with the
connected cochains $\lambda_{ni_{n}}^{p + 1} (N) \in C^{p +
1}(P(G(N)),{\bf Z}_{2}^{add})$ when all the cells
from its supports belong to the graph $G(N)$. The equation (\ref{2.28})
for the nonzero cochain $\chi^{p} \in B_{p + 1}(P(G(N)),{\bf Z}_{2}^{add})$
implies
$$
\alpha (\chi^{p}; G(M)) - \alpha (\chi^{p}; G(N)) =
\sum_{1\, \leq \, k_{1}} \, \, \sum_{\lambda_{1i_{1}}^{p + 1} (N) \, \in \, C^{p +
1}(P(G(N)),{\bf Z}_{2}^{add}): \, \, (\ref{2.25})}
$$
$$
\left(
\prod_{i_{1}\, =\, 1}^{k_{1}} ||\lambda_{1i_{1}}^{p + 1} (N)||_{J,G(N)} \right)
\left( 1 + \alpha \left( \chi^{p}, \sum_{i_{1}\, =\, 1}^{k_{1}} \lambda_{1i_{1}}^{p + 1}
(N); G(M)\right) \right)^{- 1}
$$
$$
\times \left( 1 + \alpha \left( \chi^{p},
\sum_{i_{1}\, =\, 1}^{k_{1}} \lambda_{1i_{1}}^{p + 1} (N); G(N)\right) \right)^{- 1}
\Biggl( \alpha \left( \chi^{p}, \sum_{i_{1}\, =\, 1}^{k_{1}} \lambda_{1i_{1}}^{p + 1}
(N); G(N)\right)
$$
$$
- \alpha \left( \chi^{p}, \sum_{i_{1}\, =\, 1}^{k_{1}}
\lambda_{1i_{1}}^{p + 1} (N); G(M)\right) \Biggr) +
\sum_{1\, \leq \, k_{1}} \sum_{{\lambda_{1i_{1}}^{p + 1} (M) \, \in \, C^{p + 1}(P(G(M)),{\bf
Z}_{2}^{add}),} \atop {\lambda_{1i_{1}}^{p + 1} (M)\, \notin \, C^{p +
1}(P(G(N)),{\bf Z}_{2}^{add}): \, \, (\ref{2.25})}}
$$
\begin{equation}
\label{2.47}
\left( \prod_{i_{1}\, =\, 1}^{k_{1}}
||\lambda_{1i_{1}}^{p + 1} (M)||_{J,G(M)} \right)
\left( 1 + \alpha \left( \chi^{p}, \sum_{i_{1}\, =\, 1}^{k_{1}} \lambda_{1i_{1}}^{p
+ 1} (M); G(M)\right) \right)^{- 1}.
\end{equation}
The equations (\ref{2.33}), (\ref{2.36}) imply
$$
\alpha \left( \chi^{p}, \sum_{i_{1}\, =\, 1}^{k_{1}}
\lambda_{1i_{1}}^{p + 1} (N),...,\sum_{i_{n}\, =\, 1}^{k_{n}}  \lambda_{ni_{n}}^{p +
1} (N); G(M)\right)
$$
$$
- \alpha \left( \chi^{p},
\sum_{i_{1}\, =\, 1}^{k_{1}} \lambda_{1i_{1}}^{p + 1}(N),...,\sum_{i_{n}\, =\, 1}^{k_{n}}
\lambda_{ni_{n}}^{p + 1} (N); G(N)\right) =
$$
$$
\sum_{1\, \leq \, k_{n + 1}}
\, \, \sum_{\lambda_{(n + 1)i_{n + 1}}^{p + 1} (N):\, \, (\ref{2.34}),\,
\, n \, \rightarrow \, n + 1} \left( \prod_{i_{n + 1}\, =\, 1}^{k_{n + 1}} ||
\lambda_{(n + 1)i_{n + 1}}^{p + 1} (N)||_{P(G(N))} \right)
$$
$$
\times
\left( 1 + \alpha \left( \chi^{p}, \sum_{i_{1}\, =\, 1}^{k_{1}}
\lambda_{1i_{1}}^{p + 1} (N),..., \sum_{i_{n + 1}\, =\, 1}^{k_{n + 1}} \lambda_{(n +
1)i_{n + 1}}^{p + 1} (N); G(M)\right) \right)^{- 1}
$$
$$
\times
\\ \left( 1 + \alpha \left( \chi^{p}, \sum_{i_{1}\, =\, 1}^{k_{1}} \lambda_{1i_{1}}^{p + 1}
(N),...,\sum_{i_{n + 1}\, =\, 1}^{k_{n + 1}} \lambda_{(n + 1)i_{n + 1}}^{p + 1} (N); G(N)\right)
\right)^{- 1}
$$
$$
\times
\Biggl( \alpha \left( \chi^{p}, \sum_{i_{1}\, =\, 1}^{k_{1}} \lambda_{1i_{1}}^{p +
1}(N),...,\sum_{i_{n + 1}\, =\, 1}^{k_{n + 1}} \lambda_{(n + 1)i_{n + 1}}^{p + 1} (N);
G(N)\right)
$$
$$
- \alpha \left( \chi^{p}, \sum_{i_{1}\, =\, 1}^{k_{1}}
\lambda_{1i_{1}}^{p + 1} (N),...,\sum_{i_{n + 1}\, =\, 1}^{k_{n + 1}}
\lambda_{(n + 1)i_{n + 1}}^{p + 1} (N); G(M)\right) \Biggr)
$$
$$
+
\sum_{1\, \leq \, k_{n + 1}} \, \, \sum_{{\lambda_{n + 1}^{p + 1} (M) \, \notin \, C^{p +
1}(P(G(N)),{\bf Z}_{2}^{add}):} \atop {(\ref{2.34}),\, \, n \,
\rightarrow \, n + 1}} \left( \prod_{i_{n + 1}\, =\, 1}^{k_{n + 1}} || \lambda_{(n +
1)i_{k + 1}}^{p + 1} (M)||_{P(G(M))} \right)
$$
\begin{equation}
\label{2.48} \times
\left( 1 + \alpha \left( \chi^{p}, \sum_{i_{1}\, =\, 1}^{k_{1}} \lambda_{1i_{1}}^{p
+ 1} (N),...,\sum_{i_{n}\, =\, 1}^{k_{n}} \lambda_{ni_{n}}^{p + 1}
(N), \sum_{i_{n + 1}\, =\, 1}^{k_{n + 1}} \lambda_{(n + 1)i_{n + 1}}^{p + 1}
(M); G(M)\right) \right)^{- 1}
\end{equation}
for $n = 1,...,N(\lambda, G(M))|_{M = N}$.  The relations (\ref{2.36}),
(\ref{2.37}) imply
$$
\alpha \left( \chi^{p}, \sum_{i_{1}\, =\, 1}^{k_{1}}
\lambda_{1i_{1}}^{p + 1} (N),...,
\sum_{i_{n + 1}\, =\, 1}^{k_{n + 1}} \lambda_{(n + 1)i_{n + 1}\, =\, 1}^{p + 1}
(N); G(M)\right)
$$
$$
-
\alpha \left( \chi^{p}, \sum_{i_{1}\, =\, 1}^{k_{1}} \lambda_{1i_{1}}^{p +
1}(N),...,\sum_{i_{n + 1}\, =\, 1}^{k_{n + 1}}
\lambda_{(n + 1)i_{n + 1}}^{p + 1} (N); G(N) \right) =
$$
$$
\sum_{1\, \leq \, k_{n + 2}} \, \,
\sum_{{\lambda_{(n + 2)i_{n + 2}}^{p + 1} (M) \, \notin \, C^{p + 1}(P(G(N)),{\bf Z}_{2}^{add}):}
\atop {(\ref{2.34}),\, \, n\, \rightarrow \, n + 2}}
\left( \prod_{i_{n + 2}\, =\, 1}^{k_{n + 2}} ||
\lambda_{(n + 2)i_{n + 2}}^{p + 1}
(M)||_{P(G(M))} \right)
$$
\begin{equation}
\label{2.49}
\times
\Biggl( 1 + \alpha \Biggl( \chi^{p}, \sum_{i_{1}\, =\, 1}^{k_{1}} \lambda_{1i_{1}}^{p
+ 1} (N),...,\sum_{i_{n + 1}\, =\, 1}^{k_{n + 1}}
\lambda_{(n + 1)i_{n + 1}}^{p + 1} (N), \sum_{i_{n + 2}\, =\,1}^{k_{n + 2}}
\lambda_{(n + 2)i_{n + 2}}^{p + 1} (M); G(M)\Biggr) \Biggr)^{- 1}
\end{equation}
for $n = N(\lambda, G(M))|_{M = N}$. The inequality (\ref{2.45}) and the equalities (\ref{2.47}) -
(\ref{2.49}) imply
$$
|\alpha (\chi^{p}; G(M)) - \alpha (\chi^{p}; G(N))|
\leq
\sum_{\lambda_{1}^{p + 1} (M)\, \notin \, C^{p +
1}(P(G(N)),{\bf Z}_{2}^{add}): \, \, (\ref{2.25})}
\left( \prod_{i_{1}\, =\, 1}^{k_{1}} \Bigl|
||\lambda_{1i_{1}}^{p + 1} (M)||_{J,G(M)} \Bigr| \right)  +
$$
$$
\sum_{{\lambda_{1i_{1}}^{p + 1} (N) \, \in \, C^{p +
1}(P(G(N)),{\bf Z}_{2}^{add}),}\atop
{\lambda_{2i_{2}}^{p + 1} (M) \, \notin \, C^{p +
1}(P(G(N)),{\bf Z}_{2}^{add}): \, \, (\ref{2.25}),\, (\ref{2.29})}}
\left( \prod_{i_{1}\, =\, 1}^{k_{1}} \Bigl|
||\lambda_{1i_{1}}^{p + 1} (N)||_{J,G(N)} \Bigr|
\right) \left( \prod_{i_{2}\, =\, 1}^{k_{2}} \Bigl|
||\lambda_{2i_{2}}^{p + 1} (M)||_{J,G(M)} \Bigr|
\right) + \cdots
$$
$$
+
\sum_{{\lambda_{1i_{1}}^{p + 1} (N),,...,
\lambda_{(n + 1)i_{n + 1}}^{p + 1} (N) \, \in \, C^{p +
1}(P(G(N)),{\bf Z}_{2}^{add}),}\atop
(\lambda_{(n + 2)i_{n + 2}}^{p + 1} (M) \, \notin \,
C^{p + 1}(P(G(N)),{\bf Z}_{2}^{add}):\, \, (\ref{2.25}),\, (\ref{2.34}))}
\left( \prod_{l\, =\, 1}^{n + 1}
\prod_{i_{l}\, =\,1}^{k_{l}} \Bigl| ||\lambda_{li_{l}}^{p + 1} (N)||_{J,G(N)} \Bigr| \right)
$$
\begin{equation}
\label{2.50}
\times \left( \prod_{i_{n + 2}\, =\,1}^{k_{n + 2}}\Bigl|
||\lambda_{(n + 2)i_{n + 2}}^{p + 1} (M)||_{J,G(M)} \Bigr| \right), \, \,
n = N(\lambda, G(M))|_{M = N}.
\end{equation}
For the last multiplier in the $n$ - term
($n = 1,...,N(\lambda, G(M))|_{M = N} + 2$) of the right-hand side of the inequality
(\ref{2.50}) the cochain $\lambda_{ni_{n}}^{p + 1} (M) \, \notin \, $
$C^{p + 1}(P(G(N)),{\bf Z}_{2}^{add})$. The sum
$$
\sum_{l\, =\, 1}^{n - 1} |\lambda_{li_{l}}^{p + 1} (N)|_{G(N)} +
|\lambda_{ni_{n}}^{p + 1} (M)|_{G(M)}
$$
of the cochain lengths in the $n$ - term of the right-hand side of the inequality
(\ref{2.50}) exceeds the minimal distance from the support
of the cochain $\chi^{p}$ to the boundary of the graph
$G(N)$. Now the inequalities (\ref{2.40}), (\ref{2.46}) and
(\ref{2.50}) imply that the left-hand side of the
inequality (\ref{2.50}) is small for the large graphs $G(M)$, $G(N)$:
the sequence of the correlation functions $\alpha (\chi^{p}; G(M))$
is the convergent Cauchy sequence when $G(M) \rightarrow {\bf Z}^{\times d}$.

\section{Magnetization}
\setcounter{equation}{0}

Let us consider the one-dimensional Ising model with the free
boundary conditions.  We rewrite the energy function (\ref{2.11}) in the
form (\ref{2.111}). For $2N + 1$ vertices $s_{k}^{0}$, $k = - N,...,N$,
we define the numbers
$\sigma_{k}= (- 1)^{\sigma^{0} (s_{k}^{0})} = \pm 1$ usual for
Ising model. The partition function of the Ising model
with the constant $J(s_{k}^{1}) = J$, $k = - N - 1,...,N$,
and $H(s_{k}^{0}) = H$
\begin{equation}
\label{4.2} Z_{0}(J,H;G(- N,N)) = \sum_{{\sigma_{k} \, =\, \pm \,
1,\, \, k\, =\, - N,...,N,} \atop
{\sigma_{- N - 1} \, =\, \sigma_{N + 1} \, =\, 1}} \exp \{ \beta J\sum_{k\, =\, -\,
N -\, 1}^{N} \sigma_{k} \sigma_{k + 1}
+ \beta H \sum_{k\, =\, -\, N}^{N} \sigma_{k}\}.
\end{equation}
It is possible to rewrite the definition (\ref{4.2})
\begin{equation}
\label{4.5} Z_{0}(J,H;G(- N,N)) = \hbox{Tr} \left( A^{2N}B\right)
\end{equation}
by making use of the $2\times 2$ - matrices
$$
A_{1 + \tau (\sigma_{1}), 1 + \tau (\sigma_{2})} = \exp \Biggl\{ \beta J\sigma_{1}
\sigma_{2} + \frac{\beta H}{2} (\sigma_{1} +
\sigma_{2}) \Biggr\},
$$
\begin{equation}
\label{4.4} B_{1 + \tau (\sigma_{1}), 1 + \tau (\sigma_{2})} = \exp \Biggl\{
\beta \left( J + \frac{H}{2} \right)
(\sigma_{1} + \sigma_{2}) \Biggr\},
\end{equation}
$ \sigma_{1}, \sigma_{2} = \pm 1$, the numbers $\tau (\pm 1)$ are given by the
definition (\ref{2.12}). The $2\times 2$ - matrix $A$ is
\begin{equation}
\label{4.6} A = K \left( \begin{array}{cc}

\lambda_{+} (J,H) & 0 \\

0 & \lambda_{-} (J,H)

\end{array} \right) K^{- 1},
\end{equation}
\begin{equation}
\label{4.8}
K = \left( \begin{array}{cc}

e^{\beta J}(\lambda_{+} (J,H) - \exp \{ \beta J - \beta H\})& 1 \\

1 & e^{\beta J}(\lambda_{-} (J,H) - \exp \{ \beta J + \beta H\})

\end{array} \right),
\end{equation}
$$
K^{- 1} = (e^{4\beta J}(\sinh \beta H + (\sinh^{2}
\beta H + e^{- 4\beta J})^{1/2})^{2} + 1)^{- 1}\times
$$
\begin{equation}
\label{4.9}
\left(
\begin{array}{cc}

- \, e^{\beta J}(\lambda_{-} (J,H) - \exp \{ \beta J + \beta H\})& 1 \\

 1 & - \, e^{\beta J}(\lambda_{+} (J,H) - \exp \{ \beta J - \beta H\})

\end{array} \right).
\end{equation}
The eigenvalues $\lambda_{\pm} (J,H)$ are given by the relations (\ref{1.3}). The
equalities (\ref{4.6}) - (\ref{4.9}) yield the partition function (\ref{4.5})
\begin{eqnarray}
\label{4.10}  Z_{0}(J,H;G(- N,N)) = ( - e^{4\beta J}(\sinh \beta H +
(\sinh^{2} \beta H + e^{- 4\beta J})^{1/2})^{2} - 1)^{- 1}
\Bigl\{ \lambda_{+}^{2N} (J,H) \nonumber \\ \times (\exp \{ 4\beta J + \beta H\}
(\lambda_{+} (J,H) - \exp \{ \beta J - \beta H\} )(\lambda_{-} (J,H)
- \exp \{ \beta J + \beta H\} ) \nonumber \\ + \, e^{\beta J}
((\lambda_{-} (J,H) - \exp \{ \beta J + \beta H\} ) - (\lambda_{+}
(J,H) - \exp \{ \beta J - \beta H\} )) \nonumber \\ - \exp \{ - 2\beta J - \beta H\}) +
\lambda_{-}^{2N} (J,H) (e^{ - \beta H} (\lambda_{+} (J,H) - \exp \{ \beta J - \beta H\}
) \nonumber \\ \times (\lambda_{-} (J,H) - \exp \{ \beta J + \beta H\} ) +
e^{ \beta J} (\lambda_{+} (J,H) - \exp \{ \beta J - \beta H\} )\nonumber \\ -
e^{ \beta J} (\lambda_{-} (J,H) - \exp \{ \beta J + \beta H\})
- \exp \{2\beta J + \beta H\} ) \Bigr\}.
\end{eqnarray}
For the periodic boundary conditions the matrix $B = A$ in the
relation (\ref{4.5}) and the partition function expression (\ref{1.3}) is simple.
The eigenvalues (\ref{1.3}) satisfy the inequality
$$
\label{4.11} \Biggl| \frac{\lambda_{-} (J,H)}{\lambda_{+}
(J,H)}\Biggr| = \frac{|1 - e^{- 4\beta J}|}{(\cosh \beta H +
(\sinh^{2} \beta H + e^{- 4\beta J})^{1/2})^{2}} =
$$
\begin{equation}
\label{4.11}
\frac{|1 - e^{4\beta J}|}{(e^{2\beta J}\cosh \beta H + (e^{4\beta
J}\sinh^{2} \beta H + 1)^{1/2})^{2}} < 1.
\end{equation}
By making use of the equality (\ref{4.10}) and the inequality
(\ref{4.11}) we have the same magnetization
$$
\label{4.12} \lim_{N \rightarrow \infty} (\beta (2N + 1))^{- 1}
\frac{\partial}{\partial H} \left( \ln Z_{0}(J,H;G(- N,N)) \right) =
$$
\begin{equation}
\label{4.12}
(\sinh^{2} \beta H + \exp \{ - 4\beta J\} )^{- 1/2}\sinh \beta H
\end{equation}
as the magnetization (\ref{1.5}). For the vacuum ($J = 0$) the partition function (\ref{4.2})
is
\begin{equation}
\label{4.13} Z_{0}(0,H;G(- N,N)) = (2\cosh \beta H)^{2N + 1}.
\end{equation}
Due to the relations (\ref{1.3}), (\ref{4.10}),
(\ref{4.13}) we obtain the same spontaneous magnetization
$$
\lim_{N \rightarrow \infty} \frac{\partial}{\partial x}
\left( \ln \frac{Z_{0}(J,(2N + 2)^{- 1/2}\beta^{- 1}x;G(- N,N)) }{Z_{0}(0,(2N
+ 2)^{- 1/2}\beta^{- 1}x;G(- N,N))} \right)_{x\, =\, \tanh \beta H} =
$$
\begin{equation}
\label{4.14} (\exp \{ 2\beta J\} - 1)\tanh \beta H
\end{equation}
as the spontaneous magnetization (\ref{1.8}). $2N + 2$ is the total number
of the edges of the cell complex $P(G(- N,N))$. Due to (\ref{2.14}) the two-spin
correlation function is
$$
< \sigma_{m} \sigma_{n} >_{2N + 1} \, =
(Z_{0}(J,0;G(- N,N)))^{- 1}
$$
\begin{equation}
\label{4.15}
\times \left( \sum_{{\sigma_{k} \, =\, \pm \, 1,\,
\, k\, =\, -\, N,...,N,} \atop {\sigma_{- N - 1} \, =\, \sigma_{N + 1} \, =\, 1}}
\sigma_{m} \sigma_{n} \exp \{ \beta J\sum_{k\, =\, -\, N\, -\, 1}^{N}
\sigma_{k} \sigma_{k + 1} \} \right), \, \, m,n = - N,...,N.
\end{equation}
In view of the relations (\ref{4.2}), (\ref{4.13}), (\ref{4.15})
\begin{equation}
\label{4.16} \sum_{m,n\, =\, -\, N,...,N,\, \, m\, \neq \, n} <
\sigma_{m} \sigma_{n} >_{2N + 1}  =  \beta^{- 2}
\frac{\partial^{2}}{\partial H^{2}} \left( \ln \frac{Z_{0}(J,H;G(- N,N)
}{Z_{0}(0,H;G(- N,N))} \right)_{H\, =\, 0}.
\end{equation}
The relations (\ref{1.3}), (\ref{4.10}), (\ref{4.13}), (\ref{4.16}) imply
\begin{equation}
\label{4.17} \lim_{N \rightarrow \infty} (2N + 2)^{- 1} \sum_{m,n\, =\,
-\, N,...,N,\, \, m\, \neq \, n} < \sigma_{m} \sigma_{n}
>_{2N + 1} \, = \exp \{ 2\beta J\} - 1.
\end{equation}
We choose the number $2N + 2 = \# (G(- N,N);1)$ in the left-hand side of
the equality (\ref{4.17}). It is possible to choose any number $2N + M$
for an independent of $N$ number $M$. In view of the relation (\ref{4.17})
the spontaneous magnetizations (\ref{1.12}) and (\ref{4.14}) are similar
\begin{eqnarray}
\label{4.18} \lim_{N \rightarrow \infty} \frac{\partial}{\partial x}
\left( \ln \frac{Z_{0}(J,(2N + 2)^{- 1/2}\beta^{- 1} x;G(- N,N)) }{Z_{0}(0,(2N
+ 2)^{- 1/2}\beta^{- 1} x;G(- N,N))} \right)_{x\, =\, \tanh \beta H} =
\nonumber
\\ (2\tanh \beta H) \lim_{N \rightarrow \infty} (2N + 2)^{- 1} \sum_{m,n\,
=\, -\, N,...,N,\, \, m\, < \, n} < \sigma_{m} \sigma_{n}
>_{2N + 1}.
\end{eqnarray}
Below we prove the equalities similar to the equality (\ref{4.18}) for the
$d$ - dimensional Ising model ($d = 1,2,3$) with the free boundary
conditions and with the interaction energy $J(s_{i}^{1})$ depending on the
edge $s_{i}^{1}$. We obtain also the equalities similar to the equality
(\ref{4.18}) for the $d$ - dimensional ${\bf Z}_{2}$ electrodynamics ($d = 2,3$)
with the free boundary conditions and with the interaction energy $J(s_{i}^{2})$
depending on the face $s_{i}^{2}$.

The partition function with the constant magnetic field $H(s_{i}^{p}) = H$
\begin{equation}
\label{4.19} Z_{p}(J,H;G(M)) = \sum_{\sigma^{p} \, \in \,
C^{p}(P(G(M)),{\bf Z}_{2}^{add})} \exp \{ - \beta \overline{H}_{0}
(\partial^{\ast} \sigma^{p} ) + \beta H\sum_{s_{i}^{p}\, \in \,
P(G(M))} (- 1)^{\sigma^{p} (s_{i}^{p})} \}
\end{equation}
is similar to the partition function (\ref{4.2}). For $p = 1$ it is possible
to consider the magnetic field $H(s_{i}^{1})$ depending on the edge $s_{i}^{1}$
orientation. For $H = 0$ the partition  function (\ref{4.19}) coincides with
the partition function (\ref{2.13}). By making use of the decomposition
(\ref{2.8})
$$
\exp \{ \beta H(- 1)^{\epsilon} \} = \cosh \beta H
\sum_{\chi \, \in \, {\bf Z}_{2}^{add}} (- 1)^{\chi \epsilon} (\tanh
\beta H)^{\tau ((- 1)^{\chi})}
$$
we get
$$
 \exp \{ \beta H\sum_{s_{i}^{p}\, \in \, P(G(M))} (-
1)^{\sigma^{p} (s_{i}^{p})} \} = (\cosh \beta H)^{\#
(G(M);p)}
$$
\begin{equation}
\label{4.21} \times \left( \sum_{\chi^{p} \, \in \, C^{p}(P(G(M)),{\bf Z}_{2}^{add})} (-
1)^{\langle \sigma^{p}, \chi^{p} \rangle} (\tanh \beta
H)^{|\chi^{p}|_{P(G(M))}} \right).
\end{equation}
The bilinear form
$\langle \sigma^{p}, \chi^{p} \rangle$, the mapping $\tau ((- 1)^{\chi})$
and the length $|\chi^{p}|_{P(G(M))}$ are given by the relations (\ref{2.5}),
(\ref{2.12}) and (\ref{2.38}). If the magnetic field magnetic field
$H(s_{i}^{1})$ depends on the edge $s_{i}^{1}$ orientation, the
right-hand side of the equality (\ref{4.21}) is not so simple. The
relations (\ref{2.16}), (\ref{4.19}), (\ref{4.21}) imply
\begin{eqnarray}
\label{4.221}
Z_{p}(J,H;G(M)) = Z_{p}(J,0;G(M)) (\cosh \beta H)^{\#
(G(M);p)} S_{p}(J,H;G(M)), \nonumber \\
S_{p}(J,H;G(M)) = \sum_{\chi^{p} \, \in \,
B_{p}(P(G(M)),{\bf Z}_{2}^{add})} (\tanh \beta
H)^{|\chi^{p}|_{P(G(M))}} \alpha (\chi^{p} ;G(M)).
\end{eqnarray}
The correlation function (\ref{2.16}) is equal
to zero for $\chi^{p} \notin B_{p}(P(G(M)),{\bf Z}_{2}^{add})$. The relation
(\ref{4.221}) for $p = 0$ is obtained in the paper
\cite{9}. For the vacuum ($J(s_{i}^{p + 1}) = 0$) the relations (\ref{2.11}),
(\ref{4.19}) imply
\begin{equation}
\label{4.231} Z_{p}(0,H;G(M)) = (2\cosh \beta H)^{\# (G(M);p)}.
\end{equation}
The "energy" of the constant magnetic field $H(s_{j}^{p}) = H$ for a non-boundary
cell $s_{i}^{p + 1} \in G(M)$ is the product $H^{2p + 2}$ of the
magnetic fields corresponding to $2p + 2$ boundary cells
$s_{j}^{p} \in \partial s_{i}^{p + 1}$. The total "energy" of the magnetic field
$H(s_{j}^{p}) = H$ is the sum over the cells $s_{i}^{p + 1}$
\begin{equation}
\label{4.23} \sum_{s_{i}^{p + 1}\, \in \, P(G(M))}
\left( \prod_{s_{j}^{p}\, :\, \, (s_{i}^{p + 1}:\, s_{j}^{p})\, =\, 1}
H(s_{j}^{p}) \right) \approx (\# (G(M);p + 1))H^{2p + 2}.
\end{equation}
We neglect the boundary cells $s_{i}^{p + 1}$ from $P(G(M))$. The "re-normalized"
magnetic field
\begin{equation}
\label{4.232}
H(s_{j}^{p}) = (\# (G(M);p + 1))^{- 1/(2p\, + 2)}\beta^{- 1}\tanh
\beta H
\end{equation}
yields the constant "re-normalized total energy" (\ref{4.23}). In view of the
relations (\ref{4.221}), (\ref{4.231}) we get the spontaneous
magnetization for the "re-normalized" magnetic field (\ref{4.232})
$$
\lim_{G(M) \rightarrow {\bf Z}^{\times d}}
\frac{\partial}{\partial x} \left( \ln \frac{Z_{p}(J,(\# (G(M);p + 1))^{-
1/(2p\, + 2)} \beta^{- 1} x;G(M)) }{Z_{p}(0,(\# (G(M);p + 1))^{-
1/(2p\, + 2)} \beta^{- 1} x;G(M))} \right)_{x\, =\, \tanh
\beta H} =
$$
\begin{equation}
\label{4.24}
\lim_{G(M) \rightarrow {\bf Z}^{\times d}}
\frac{\partial}{\partial x} \left( \ln S_{p}(J,(\# (G(M);p + 1))^{-
1/(2p\, + 2)} \beta^{- 1} x;G(M))\right)_{x\, =\, \tanh \beta H}.
\end{equation}
Let us introduce the set of the connected cochains
$\lambda_{1i_{1}}^{p + 1} (M) \in C^{p + 1}(P(G(M)),{\bf Z}_{2}^{add})$:
$$
\partial \lambda_{1i_{1}}^{p + 1} (M) \neq 0;
$$
\begin{equation}
\label{4.28}
(s_{i}^{p + 1}:s_{l}^{p}) (s_{j}^{p + 1}:s_{l}^{p}) = 0, \, \,
\lambda_{1i_{1}}^{p + 1} (M) (s_{i}^{p + 1}) = 1,\, \,
\lambda_{1j_{1}}^{p + 1} (M) (s_{j}^{p + 1}) = 1,\, \, i_{1} \neq j_{1},
\end{equation}
$i_{1},j_{1} = 1,...,k_{1}$. The integer $k_{1} \leq \# (G(M);p + 1)$.
The relations (\ref{2.28}), (\ref{4.221}) imply
$$
S_{p}(J,(\# (G(M);p + 1))^{- 1/(2p\, + 2)} \beta^{-
1} \tanh (\beta H);G(M)) =
$$
$$
1 + \sum_{1\, \leq \, k_{1}} \, \, \sum_{\lambda_{1i_{1}}^{p + 1} (M) \,
: \, \, (\ref{4.28})}
\left( \prod_{i_{1}\, =\, 1}^{k_{1}}
||\lambda_{1i_{1}}^{p + 1} (M)||_{J,G(M)}
\right)
$$
$$
\times \left( \tanh ((\# (G(M);p + 1))^{-
1/(2p\, + 2)} \tanh (\beta H))\right)^{\sum_{i_{1}\, =\, 1}^{k_{1}} |\partial \lambda_{1i_{1}}^{p +
1} (M)|_{P(G(M))}}
$$
\begin{equation}
\label{4.27}
\times \left( 1 + \alpha \left( \sum_{i_{1}\, =\, 1}^{k_{1}} \partial
\lambda_{1i_{1}}^{p + 1} (M) ,\sum_{i_{1}\, =\, 1}^{k_{1}} \lambda_{1i_{1}}^{p + 1}
(M); G(M)\right) \right)^{- 1}.
\end{equation}
Due to the definition (\ref{2.14}) the correlation function $\alpha
(0; G(M)) = 1$. It is easy to verify
$$
\frac{d}{dx} \tanh x  = (\cosh x)^{- 2}, \, \, |\tanh
x| \leq |x|,
$$
\begin{equation}
\label{4.31}
x = ( \# (G(M);p + 1))^{- 1/(2p + 2)} \tanh (\beta H).
\end{equation}
The inequality (\ref{2.45}) implies
\begin{equation}
\label{4.20}
\left( 1 + \alpha \left( \sum_{i_{1}\, =\, 1}^{k_{1}} \partial
\lambda_{1i_{1}}^{p + 1} (M) ,\sum_{i_{1}\, =\, 1}^{k_{1}} \lambda_{1i_{}}^{p + 1}
(M); G(M)\right) \right)^{- 1} \leq 1.
\end{equation}
If the support of the connected cochain
$\mu^{p + 1} \in C^{p + 1}(P(G(M)),{\bf Z}_{2}^{add})$ consists of the only
cell $s_{i}^{p + 1}$, then  the length of its boundary
\begin{equation}
\label{4.321}
|\partial \mu^{p + 1} |_{P(G(M))} = 2p + 2.
\end{equation}
If the connected cochain
$\mu^{p + 1} \in C^{p + 1}(P(G(M)),{\bf Z}_{2}^{add})$ satisfies the equation
(\ref{4.321}), the length $|\mu^{p + 1} |_{P(G(M))} = 1$ for $p = 1$, $d = 2$.
The length of the cochain $|\mu^{p + 1} |_{P(G(M))}$ satisfying the equation
(\ref{4.321}) may be practically arbitrary for $p = 0$,
$d = 1,2,3$ and for $p = 1$, $d = 3$. Let us introduce the connected cochains
$\mu_{1i_{1}}^{p + 1} (M) \in C^{p + 1}(P(G(M)),{\bf Z}_{2}^{add})$,
$i_{1} = 1,...,k_{1}$, satisfying the equations
$$
|\partial \mu_{1i_{1}}^{p + 1} (M)|_{P(G(M))} = 2p + 2,
\, \, i_{1} = 1,...,k_{1};
$$
\begin{equation}
\label{4.32}
(s_{i}^{p + 1}:s_{l}^{p}) (s_{j}^{p + 1}:s_{l}^{p}) = 0, \, \,
\mu_{1i_{1}}^{p + 1} (M) (s_{i}^{p + 1}) = 1,\, \,
\mu_{1j_{1}}^{p + 1} (M) (s_{j}^{p + 1}) = 1,\, \, i_{1} \neq j_{1},
\end{equation}
$i_{1},j_{1} = 1,...,k_{1}$.

The  ratio of the total number of the shifts of the connected cochain
$\lambda^{p + 1} (M)$ from the group $C^{p + 1}(P(G(M)),{\bf Z}_{2}^{add})$
in the graph $G(M)$ and of the number $\# (G(M);p + 1)$ tends to one when
$G(M) \rightarrow {\bf Z}^{\times d}$. Let the sign of the interaction energy
$J(s_{i}^{p + 1}$) be independent of the cell $s_{i}^{p + 1}$ and the
interaction energy $J(s_{i}^{p + 1})$ satisfy the inequality (\ref{2.46}).
By making use of the inequalities (\ref{2.40}), (\ref{4.31}), (\ref{4.20})
it is possible to prove that in the right-hand side of the equality (\ref{4.27})
the terms with the cochains (\ref{4.32}) only may be nonzero when
$G(M) \rightarrow {\bf Z}^{\times d}$
$$
\lim_{G(M) \rightarrow {\bf Z}^{\times d}}
S_{p}(J,(\# (G(M);p + 1))^{- 1/(2p\, + 2)} \beta^{-
1} \tanh (\beta H);G(M)) =
$$
$$
1 + \lim_{G(M) \rightarrow {\bf Z}^{\times d}} \sum_{1\, \leq \, k_{1}}
\, \, \sum_{\mu_{1i_{1}}^{p + 1} (M): \, \, (\ref{4.32})}
\left( \prod_{i_{1}\, =\, 1}^{k_{1}}
||\mu_{1i_{1}}^{p + 1} (M)||_{J,G(M)}
\right)
$$
$$
\times \left( \tanh ((\# (G(M);p + 1))^{-
1/(2p\, + 2)} \tanh (\beta H))\right)^{2k_{1}(p + 1)}
$$
\begin{equation}
\label{4.33}
\times \left( 1 + \alpha \left( \sum_{i_{1}\, =\, 1}^{k_{1}} \partial
\mu_{1i_{1}}^{p + 1} (M) ,\sum_{i_{1}\, =\, 1}^{k_{1}} \mu_{1i_{}}^{p + 1}
(M); G(M)\right) \right)^{- 1}.
\end{equation}
Due to the relations (\ref{2.1612}), (\ref{2.1613}), (\ref{2.1622}),
(\ref{2.1623}) the left-hand side of the inequality (\ref{4.20}) is
equal to one for $p = 0$, $d = 1$ and $p = 1$, $d = 2$. For these
theories the proof of the relation (\ref{4.58}) similar to (\ref{4.18})
is continued from the relation (\ref{4.56}). The correlation function
$$
\alpha \left( \sum_{i_{1}\, =\, 1}^{k_{1}} \partial
\mu_{1i_{1}}^{p + 1} (M) ,\sum_{i_{1}\, =\, 1}^{k_{1}} \mu_{1i_{1}}^{p + 1} (M);
G(M)\right)
$$
in the right-hand side of the equality (\ref{4.33})
satisfies the equation (\ref{2.33}). The connected cochains
$\lambda_{2i_{2}}^{p + 1} (M)$, $i_{2} = 1,...,k_{2}$, in the equality
(\ref{2.33}) satisfy the equations (\ref{2.29}) for the cochains
$\mu_{1i_{1}}^{p + 1} (M)$ instead of the cochains $\lambda_{1i_{1}}^{p + 1} (M)$.
We divide the cochains $\lambda_{2i_{2}}^{p + 1} (M)$, $i_{2} = 1,...,k_{2}$,
into two sets. The first set consists of the connected cochains
$\mu_{2i_{2}}^{p + 1} (M) \in $ $C^{p + 1}(P(G(M)),{\bf Z}_{2}^{add})$,
$i_{2} = 1,...,k_{2}$, satisfying the equations: for every number
$i_{2} = 1,...,k_{2}$ there is only one number $i_{1} = 1,...,k_{1}$ such that
\begin{equation}
\label{4.35}
\exists \, i,j,l, \, \, (s_{i}^{p + 1}:s_{l}^{p}) (s_{j}^{p + 1}:s_{l}^{p}) = 1,\, \,
\mu_{2i_{2}}^{p + 1} (M) (s_{i}^{p + 1}) = 1, \, \,
\mu_{1i_{1}}^{p + 1} (M) (s_{j}^{p + 1}) = 1.
\end{equation}
The connected cochains
$\mu_{2i_{2}}^{p + 1} (M) \in $ $C^{p + 1}(P(G(M)),{\bf Z}_{2}^{add})$,
$i_{2} = 1,...,k_{2}$, satisfy also the equations similar to the first and the
third equations (\ref{2.29})
$$
\partial \mu_{2i_{2}}^{p + 1} (M) = 0;
$$
\begin{equation}
\label{4.36}
(s_{i}^{p + 1}:s_{l}^{p}) (s_{j}^{p + 1}:s_{l}^{p}) = 0,
\, \, \mu_{2i_{2}}^{p + 1} (M) (s_{i}^{p + 1}) = 1,\, \,
\mu_{2j_{2}}^{p + 1} (M) (s_{j}^{p + 1}) = 1,\, \, i_{2} \neq j_{2},
\end{equation}
$i_{2},j_{2} = 1,...,k_{2}$.

The second set consists of the connected cochains
$\nu_{2i_{2}}^{p + 1} (M) \in$$C^{p + 1}(P(G(M)),{\bf Z}_{2}^{add})$,
$i_{2} = 1,...,k_{2}$ satisfying the equations: for every number
$i_{2} = 1,...,k_{2}$ there exists the number $i_{1} = 1,...,k_{1}$ such that
\begin{equation}
\label{4.361}
\exists \, i,j,l, \, \, (s_{i}^{p + 1}:s_{l}^{p}) (s_{j}^{p + 1}:s_{l}^{p}) = 1,\, \,
\nu_{2i_{2}}^{p + 1} (M) (s_{i}^{p + 1}) = 1, \, \,
\mu_{1i_{1}}^{p + 1} (M) (s_{j}^{p + 1}) = 1
\end{equation}
and there exists the number $i_{2} = 1,...,k_{2}$ such that
$$
\exists \, i,j,l, \, \, (s_{i}^{p + 1}:s_{l}^{p}) (s_{j}^{p + 1}:s_{l}^{p}) = 1,\, \,
\nu_{2i_{2}}^{p + 1} (M) (s_{i}^{p + 1}) = 1, \, \,
\mu_{1i_{1}}^{p + 1} (M) (s_{j}^{p + 1}) = 1,
$$
\begin{equation}
\label{4.37}
\exists \, i,j,l, \, \, (s_{i}^{p + 1}:s_{l}^{p}) (s_{j}^{p + 1}:s_{l}^{p}) = 1,\, \,
\nu_{2i_{2}}^{p + 1} (M) (s_{i}^{p + 1}) = 1, \, \,
\mu_{1j_{1}}^{p + 1} (M) (s_{j}^{p + 1}) = 1
\end{equation}
for at least two different numbers $i_{1},j_{1} = 1,...,k_{1}$. The connected cochains
$\nu_{2i_{2}}^{p + 1} (M) \in $ $C^{p + 1}(P(G(M)),{\bf Z}_{2}^{add})$,
$i_{2} = 1,...,k_{2}$, satisfy also the equations similar to the equations (\ref{4.36})
$$
\partial \nu_{2i_{2}}^{p + 1} (M) = 0;
$$
\begin{equation}
\label{4.38} (s_{i}^{p + 1}:s_{l}^{p}) (s_{j}^{p + 1}:s_{l}^{p}) = 0,
\, \, \nu_{2i_{2}}^{p + 1} (M) (s_{i}^{p + 1}) = 1,\, \,
\nu_{2j_{2}}^{p + 1} (M) (s_{j}^{p + 1}) = 1,\, \, i_{2} \neq j_{2},
\end{equation}
$i_{2},j_{2} = 1,...,k_{2}$. We divide the sum (\ref{2.33}) into two parts
$$
\alpha_{2} \left( \sum_{i_{1}\, =\, 1}^{k_{1}} \partial
\mu_{1i_{1}}^{p + 1} (M) ,\sum_{i_{1}\, =\, 1}^{k_{1}} \mu_{1i_{1}}^{p + 1} (M);
G(M)\right) =
$$
$$
\sum_{1\, \leq \, k_{2}} \, \,
\sum_{\mu_{2i_{2}}^{p + 1} (M):\, \, (\ref{4.35}),\, \, (\ref{4.36})} \left(
\prod_{i_{2}\, =\, 1}^{k_{2}}
||\mu_{2i_{2}}^{p + 1} (M) ||_{J,G(M)} \right)
$$
\begin{equation}
\label{4.41}
\times \left( 1 + \alpha \left( \sum_{i_{1}\, =\, 1}^{k_{1}} \partial
\mu_{1i_{1}}^{p + 1} (M) ,\sum_{i_{1}\, =\, 1}^{k_{1}} \mu_{1i_{1}}^{p + 1} (M),
\sum_{i_{2}\, =\, 1}^{k_{2}} \mu_{2i_{2}}^{p + 1} (M); G(M)\right)
\right)^{- 1},
\end{equation}
$$
\beta_{2} \left( \sum_{i_{1}\, =\, 1}^{k_{1}} \partial
\mu_{1i_{1}}^{p + 1} (M) ,\sum_{i_{1}\, =\, 1}^{k_{1}} \mu_{1i_{1}}^{p + 1} (M);
G(M)\right) =
$$
$$
\sum_{1\, \leq \, k_{2}} \, \,
\sum_{\nu_{2i_{2}}^{p + 1} (M):\, \, (\ref{4.361}) - (\ref{4.38})} \left(
\prod_{i_{2}\, =\, 1}^{k_{2}}
||\nu_{2i_{2}}^{p + 1} (M) ||_{J,G(M)} \right)
$$
\begin{equation}
\label{4.42}
\times \left( 1 + \alpha \left( \sum_{i_{1}\, =\, 1}^{k_{1}} \partial
\mu_{1i_{1}}^{p + 1} (M) ,\sum_{i_{1}\, =\, 1}^{k_{1}} \mu_{1i_{1}}^{p + 1} (M),
\sum_{i_{2}\, =\, 1}^{k_{2}} \nu_{2i_{2}}^{p + 1} (M); G(M)\right)
\right)^{- 1}.
\end{equation}
If the sign of the interaction energy $J(s_{i}^{p + 1}$) is
independent of the cell $s_{i}^{p + 1}$ and the interaction energy
$J(s_{i}^{p + 1})$ satisfies the inequality (\ref{2.46}), then the
inequalities (\ref{2.44}), (\ref{2.45}), (\ref{2.40})) and the
equalities (\ref{4.41}), (\ref{4.42}) imply
$$
0 \leq \alpha_{2} \left( \sum_{i_{1}\, =\, 1}^{k_{1}}
\partial \mu_{1i_{1}}^{p + 1} (M) ,\sum_{i_{1}\, =\, 1}^{k_{1}} \mu_{1i_{1}}^{p
+ 1} (M); G(M)\right) < 1,
$$
\begin{equation}
\label{4.43}
0 \leq \beta_{2} \left( \sum_{i_{1}\, =\, 1}^{k_{1}}
\partial \mu_{1i_{1}}^{p + 1} (M) ,\sum_{i_{1}\, =\, 1}^{k_{1}} \mu_{1i_{1}}^{p
+ 1} (M); G(M)\right) < 1,
\end{equation}
\begin{eqnarray}
\label{4.44} \left( 1 + \alpha \left( \sum_{i_{1}\, =\, 1}^{k_{1}}
\partial \mu_{1i_{1}}^{p + 1} (M) ,\sum_{i_{1}\, =\, 1}^{k_{1}} \mu_{1i_{1}}^{p
+ 1} (M);
G(M)\right) \right)^{- 1} = \nonumber \\
\sum_{m\, =\, 0}^{\infty} \left( 1 + \alpha_{2} \left(
\sum_{i_{1}\, =\, 1}^{k_{1}} \partial \mu_{1i_{1}}^{p + 1} (M) ,\sum_{i_{1}\,
=\, 1}^{k_{1}}
\mu_{1i_{1}}^{p + 1} (M); G(M)\right) \right)^{- m - 1} \nonumber \\
\times \left( - \beta_{2} \left( \sum_{i_{1}\, =\, 1}^{k_{1}} \partial
\mu_{1i_{1}}^{p + 1} (M), \sum_{i_{1}\, =\, 1}^{k_{1}} \mu_{1i_{1}}^{p + 1} (M);
G(M)\right) \right)^{m}.
\end{eqnarray}
Let us substitute the equality (\ref{4.44}) into the right-hand side
of the equality (\ref{4.33}). Now every term of the sum (\ref{4.33}) with
the term of the sum (\ref{4.44}) for $m \geq 1$ contains the cochain
$\nu_{2i_{2}}^{p + 1} (M)$ connecting together due to the relations
(\ref{4.37}) at least two cochains
$\mu_{1j_{1}}^{p + 1} (M)$, $\mu_{1l_{1}}^{p + 1} (M)$ in
the sum (\ref{4.33}). These connected together cochains can move on
the graph $G(M)$ as one connected cochain. Hence the inequality
(\ref{4.31}) implies
\begin{eqnarray}
\label{4.45}  \lim_{G(M) \rightarrow {\bf Z}^{\times d}} S_{p}(J,(\#
(G(M);p + 1))^{- 1/(2p\, + 2)} \beta^{- 1} \tanh (\beta H);G(M)) =
\nonumber \\ 1 + \lim_{G(M) \rightarrow {\bf Z}^{\times d}} \sum_{1\, \leq \, k_{1}}
\, \, \sum_{\mu_{1i_{1}}^{p + 1} (M): \, \, (\ref{4.32})} \left(
\prod_{i_{1}\, =\, 1}^{k_{1}} ||\mu_{1i_{1}}^{p + 1}
(M)||_{J,G(M)} \right) \nonumber \\ \times \left( \tanh ((\# (G(M);p + 1))^{- 1/(2p\, +
2)} \tanh (\beta H))\right)^{2k_{1}(p + 1)} \nonumber
\\
\times \left( 1 + \alpha_{2} \left( \sum_{i_{1}\, =\, 1}^{k_{1}} \partial
\mu_{1i_{1}}^{p + 1} (M) ,\sum_{i_{1}\, =\, 1}^{k_{1}}
\mu_{1i_{1}}^{p + 1} (M);G(M)\right) \right)^{- 1}.
\end{eqnarray}
We define the set of the cochains
$\mu_{ni_{n}}^{p + 1} (M)$, $i_{n} = 1,...,k_{n}$, $n = 1,2,...$,
from the group $C^{p + 1}(P(G(M)),{\bf Z}_{2}^{add})$. The cochains
$\mu_{1i_{1}}^{p + 1} (M)$, $i_{1} = 1,...,k_{1}$, satisfy the
equations (\ref{4.32}). The cochains
$\mu_{ni_{n}}^{p + 1} (M)$, $i_{n} = 1,...,k_{n}$, for $n \geq 2$ satisfy the
equations: for every number $i_{n}$ there exists the sequence of the cochains
$\mu_{1i_{1}}^{p + 1} (M)$,...,$\mu_{ni_{n}}^{p + 1} (M)$ such that
\begin{equation}
\label{4.46}
\exists \, i,j,l, \, \, (s_{i}^{p + 1}:s_{l}^{p}) (s_{j}^{p + 1}:s_{l}^{p}) = 1,\, \,
\mu_{mi_{m}}^{p + 1} (M) (s_{i}^{p + 1}) = 1, \, \,
\mu_{(m + 1)i_{m + 1}}^{p + 1} (M) (s_{j}^{p + 1}) = 1,
\end{equation}
$m = 1,...,n - 1$, and any two sequences (\ref{4.46})
$$
\exists \, i,j,l, \, \, (s_{i}^{p + 1}:s_{l}^{p}) (s_{j}^{p + 1}:s_{l}^{p}) = 1,\, \,
\mu_{mi_{m}}^{p + 1} (M) (s_{i}^{p + 1}) = 1, \, \,
\mu_{(m + 1)i_{m + 1}}^{p + 1} (M) (s_{j}^{p + 1}) = 1,
$$
\begin{equation}
\label{4.47}
\exists \, i,j,l, \, \, (s_{i}^{p + 1}:s_{l}^{p}) (s_{j}^{p + 1}:s_{l}^{p}) = 1,\, \,
\mu_{mj_{m}}^{p + 1} (M) (s_{i}^{p + 1}) = 1, \, \,
\mu_{(m + 1)j_{m + 1}}^{p + 1} (M) (s_{j}^{p + 1}) = 1,
\end{equation}
$m = 1,...,n - 1$, with the same end: $i_{n} = j_{n}$ have the same beginning: $i_{1} = j_{1}$.
The cochains $\mu_{ni_{n}}^{p + 1} (M)$, $i_{n} = 1,...,k_{n}$, for $n \geq 2$
satisfy also the equations similar to the equations (\ref{2.34})
$$
\partial \mu_{ni_{n}}^{p + 1} (M) = 0,\, \, i_{n} = 1,...,k_{n};
$$
$$
(s_{i}^{p + 1}:s_{l}^{p}) (s_{j}^{p + 1}:s_{l}^{p}) = 0,\, \,
\mu_{mi_{m}}^{p + 1} (M) (s_{i}^{p + 1}) = 1, \, \,
\mu_{ni_{n}}^{p + 1} (M) (s_{j}^{p + 1}) = 1,
$$
$$
i_{m} = 1,...,k_{m},\, \, i_{n} = 1,...,k_{n},\, \, m = 1,...,n - 2 \geq 1;
$$
\begin{equation}
\label{4.48}
(s_{i}^{p + 1}:s_{l}^{p}) (s_{j}^{p + 1}:s_{l}^{p}) = 0,\, \,
\mu_{ni_{n}}^{p + 1} (M) (s_{i}^{p + 1}) = 1, \, \,
\mu_{nj_{n}}^{p + 1} (M) (s_{j}^{p + 1}) = 1, \, \, i_{n} \neq j_{n},
\end{equation}
$i_{n},j_{n} = 1,...,k_{n}$. For $n = 2$ the equations (\ref{4.46}) - (\ref{4.48})
coincide with the equations (\ref{4.35}), (\ref{4.36}). The equations (\ref{4.47})
mean that the set of the the cochains $\mu_{li_{l}}^{p + 1} (M) \in$
$C^{p + 1}(P(G(M)),{\bf Z}_{2}^{add})$, $i_{l} = 1,...,k_{l}$,
$l = 1,2,...$, is  $k_{1}$ cochain trees with the trunks
$\mu_{11}^{p + 1} (M),...,$

\noindent $\mu_{1k_{1}}^{p + 1} (M)$. By repeating
the proof of the equality (\ref{4.45}) it is possible to prove that the
correlation function
$$
\alpha_{2} \left( \sum_{i_{1}\,
=\, 1}^{k_{1}} \partial \mu_{1i_{1}}^{p + 1} (M) ,\sum_{i_{1}\, =\,
1}^{k_{1}} \mu_{1i_{1}}^{p + 1} (M); G(M)\right)
$$
in the right-hand side of the equality (\ref{4.45}) may be considered
as the first term of the sequence of the correlation functions
\begin{eqnarray}
\label{4.49} \alpha_{n} \left( \sum_{i_{1}\, =\, 1}^{k_{1}} \partial
\mu_{1i_{1}}^{p + 1} (M) ,\sum_{i_{1}\, =\, 1}^{k_{1}} \mu_{1i_{1}}^{p + 1}
(M),...,\sum_{i_{n - 1}\, =\, 1}^{k_{n - 1}} \mu_{(n - 1)i_{n - 1}}^{p + 1} (M);
G(M)\right) = \nonumber \\ \sum_{1\, \leq \, k_{n}}
\, \, \sum_{\mu_{ni_{n}}^{p + 1} (M):\, \, (\ref{4.46}) - (\ref{4.48})}
\left( \prod_{i_{n}\, =\, 1}^{k_{n}}
||\mu_{ni_{n}}^{p + 1} (M) ||_{J,G(M)} \right) \nonumber \\
\times \left( 1 + \alpha_{n + 1} \left( \sum_{i_{1}\, =\, 1}^{k_{1}} \partial
\mu_{1i_{1}}^{p + 1} (M) ,\sum_{i_{1}\, =\, 1}^{k_{1}} \mu_{1i_{1}}^{p + 1}
(M),..., \sum_{i_{n}\, =\, 1}^{k_{n}} \mu_{ni_{n}}^{p + 1} (M); G(M) \right)
\right)^{- 1},
\end{eqnarray}
$n = 2,...,N - 1$. For $n = N$ the correlation function
\begin{eqnarray}
\label{4.50} \alpha_{N} \left( \sum_{i_{1}\, =\, 1}^{k_{1}} \partial
\mu_{1i_{1}}^{p + 1} (M) ,\sum_{i_{1}\, =\, 1}^{k_{1}} \mu_{1i_{1}}^{p + 1}
(M),...,\sum_{i_{N - 1}\, =\, 1}^{k_{N - 1}} \mu_{(N - 1)i_{N - 1}}^{p + 1} (M);
G(M)\right) = \nonumber \\ \sum_{1\, \leq \, k_{N}}
\, \, \sum_{\mu_{Ni_{N}}^{p + 1} (M):\, \, (\ref{4.46}) -
(\ref{4.48}),\, \, n\, \rightarrow \, N} \left( \prod_{i_{N}\, =\, 1}^{k_{N}}
||\mu_{Ni_{N}}^{p + 1} (M) ||_{J,G(M)} \right) \nonumber \\
\times \left( 1 + \alpha \left( \sum_{i_{1}\, =\, 1}^{k_{1}} \partial
\mu_{1i_{1}}^{p + 1} (M) ,\sum_{i_{1}\, =\, 1}^{k_{1}} \mu_{1i_{1}}^{p + 1}
(M),..., \sum_{i_{N}\, =\, 1}^{k_{N}} \mu_{Ni_{N}}^{p + 1} (M); G(M) \right)
\right)^{- 1}.
\end{eqnarray}
$N$ is an arbitrary integer independent of the graph $G(M)$. For $N = 2$ the
relation (\ref{4.50}) coincides with the relation (\ref{4.41}).

Let us define the sequence of the correlation functions
$$
\alpha_{n}^{(N)} \left( \sum_{i_{1}\, =\, 1}^{k_{1}}
\partial \mu_{1i_{1}}^{p + 1} (M) ,\sum_{i_{1}\, =\, 1}^{k_{1}} \mu_{1i_{1}}^{p
+ 1} (M),...,\sum_{i_{n - 1}\, =\, 1}^{k_{n - 1}} \mu_{(n - 1)i_{n - 1}}^{p + 1}
(M); G(M)\right),
$$
$ n = 2,...,N - 1$, satisfying the relations (\ref{4.49}) where the correlation function
\begin{equation}
\label{4.51} \alpha_{N}^{(N)} \left( \sum_{i_{1}\, =\, 1}^{k_{1}}
\partial \mu_{1i_{1}}^{p + 1} (M) ,\sum_{i_{1}\, =\, 1}^{k_{1}} \mu_{1i_{1}}^{p
+ 1} (M),...,\sum_{i_{N - 1}\, =\, 1}^{k_{N - 1}} \mu_{(N - 1)i_{N - 1}}^{p + 1}
(M); G(M)\right) = 0
\end{equation}
instead of the correlation function (\ref{4.50}). The relations (\ref{4.49}) -
(\ref{4.51}) imply
\begin{eqnarray}
\label{4.52} \alpha_{n}^{(N)} \left( \sum_{i_{1}\, =\, 1}^{k_{1}}
\partial \mu_{1i_{1}}^{p + 1} (M) ,\sum_{i_{1}\, =\, 1}^{k_{1}} \mu_{1i_{1}}^{p
+ 1} (M),...,\sum_{i_{n - 1}\, =\, 1}^{k_{n - 1}} \mu_{(n - 1)i_{n - 1}}^{p + 1}
(M); G(M)\right) \nonumber \\ - \alpha_{n} \left( \sum_{i_{1}\, =\,
1}^{k_{1}} \partial \mu_{1i_{1}}^{p + 1} (M) ,\sum_{i_{1}\, =\, 1}^{k_{1}} \mu_{1i_{1}}^{p
+ 1} (M),...,\sum_{i_{n - 1}\, =\, 1}^{k_{n - 1}} \mu_{(n - 1)i_{n - 1}}^{p + 1}
(M); G(M)\right) = \nonumber \\ \sum_{1\, \leq \, k_{n}}
\, \, \sum_{\mu_{ni_{n}}^{p + 1} (M):\, \, (\ref{4.46}) - (\ref{4.48})}
\left( \prod_{i_{n}\, =\, 1}^{k_{n}}
||\mu_{ni_{n}}^{p + 1} (M) ||_{J,G(M)} \right) \nonumber \\
\times \left( 1 + \alpha_{n + 1} \left( \sum_{i_{1}\, =\, 1}^{k_{1}} \partial
\mu_{1i_{1}}^{p + 1} (M) ,\sum_{i_{1}\, =\, 1}^{k_{1}} \mu_{1i_{1}}^{p + 1}
(M),..., \sum_{i_{n}\, =\, 1}^{k_{n}} \mu_{ni_{n}}^{p + 1} (M); G(M) \right)
\right)^{- 1} \nonumber \\
\times \left( 1 + \alpha_{n + 1}^{N} \left( \sum_{i_{1}\, =\, 1}^{k_{1}}
\partial \mu_{1i_{1}}^{p + 1} (M) ,\sum_{i_{1}\, =\, 1}^{k_{1}} \mu_{1i_{1}}^{p
+ 1} (M),..., \sum_{i\, =\, 1}^{k_{n}} \mu_{ni}^{p + 1} (M); G(M)
\right) \right)^{- 1} \nonumber \\
\times \Biggl( \alpha_{n + 1} \left( \sum_{i_{1}\, =\, 1}^{k_{1}}
\partial \mu_{1i_{1}}^{p + 1} (M) ,\sum_{i_{1}\, =\, 1}^{k_{1}} \mu_{1i_{1}}^{p
+ 1} (M),...,\sum_{i_{n}\, =\, 1}^{k_{n}} \mu_{ni_{n}}^{p + 1}
(M); G(M)\right) \nonumber \\
- \alpha_{n + 1}^{(N)} \left( \sum_{i_{1}\, =\, 1}^{k_{1}}
\partial \mu_{1i_{1}}^{p + 1} (M) ,\sum_{i_{1}\, =\, 1}^{k_{1}} \mu_{1i_{1}}^{p
+ 1} (M),...,\sum_{i_{n}\, =\, 1}^{k_{n}} \mu_{ni_{n}}^{p + 1} (M);
G(M)\right) \Biggr),
\end{eqnarray}
\begin{eqnarray}
\label{4.53} \alpha_{N}^{(N)} \left( \sum_{i_{1}\, =\, 1}^{k_{1}}
\partial \mu_{1i_{1}}^{p + 1} (M) ,\sum_{i_{1}\, =\, 1}^{k_{1}} \mu_{1i_{1}}^{p
+ 1} (M),...,\sum_{i_{N - 1}\, =\, 1}^{k_{N - 1}} \mu_{(N - 1)i_{N - 1}}^{p + 1}
(M); G(M)\right) \nonumber \\ - \alpha_{N} \left( \sum_{i_{1}\, =\, 1}^{k_{1}}
\partial \mu_{1i_{1}}^{p + 1} (M) ,\sum_{i_{1}\, =\, 1}^{k_{1}} \mu_{1i_{1}}^{p
+ 1} (M),...,\sum_{i_{N - 1}\, =\, 1}^{k_{N - 1}} \mu_{(N - 1)i_{N - 1}}^{p + 1}
(M); G(M)\right) = \nonumber \\ - \alpha_{N} \left( \sum_{i_{1}\, =\,
1}^{k_{1}} \partial \mu_{1i_{1}}^{p + 1} (M) ,\sum_{i_{1}\, =\, 1}^{k_{1}}
\mu_{1i_{1}}^{p + 1} (M),...,\sum_{i_{N - 1}\, =\, 1}^{k_{N - 1}} \mu_{(N -
1)i_{N - 1}}^{p + 1} (M); G(M)\right).
\end{eqnarray}
If the sign of the interaction energy $J(s_{i}^{p + 1}$) is
independent of the cell $s_{i}^{p + 1}$, then the inequality
(\ref{2.44}) and the definitions (\ref{4.49}) - (\ref{4.51}) imply
\begin{eqnarray}
\label{4.54} \alpha_{n}^{(N)} \left( \sum_{i_{1}\, =\, 1}^{k_{1}}
\partial \mu_{1i_{1}}^{p + 1} (M) ,\sum_{i_{1}\, =\, 1}^{k_{1}} \mu_{1i_{1}}^{p
+ 1} (M),...,\sum_{i_{n - 1}\, =\, 1}^{k_{n - 1}} \mu_{(n - 1)i_{n - 1}}^{p + 1}
(M); G(M)\right) \geq 0, \nonumber \\ \alpha_{n} \left( \sum_{i_{1}\, =\, 1}^{k_{1}}
\partial \mu_{1i_{1}}^{p + 1} (M) ,\sum_{i_{1}\, =\, 1}^{k_{1}} \mu_{1i_{1}}^{p
+ 1} (M),...,\sum_{i_{n - 1}\, =\, 1}^{k_{n - 1}} \mu_{(n - 1)i_{n - 1}}^{p + 1}
(M); G(M)\right) \geq 0,
\end{eqnarray}
$n = 2,...,N$. If the sign of the interaction energy $J(s_{i}^{p +
1}$) is independent of the cell $s_{i}^{p + 1}$ and the interaction
energy $J(s_{i}^{p + 1})$ satisfies the inequality (\ref{2.46}),
then the inequalities (\ref{2.45}), (\ref{2.40}), (\ref{4.54}) and
the equalities (\ref{4.52}), (\ref{4.53})  imply that
the difference (\ref{4.52}), $n = 2$ is small for the large numbers
$N$. Hence we get
$$
\lim_{G(M) \rightarrow {\bf Z}^{\times d}} S_{p}(J,(\#
(G(M);p + 1))^{- 1/(2p\, + 2)} \beta^{- 1} \tanh (\beta H);G(M)) =
$$
$$
1 + \lim_{N \rightarrow \infty} \lim_{G(M) \rightarrow
{\bf Z}^{\times d}} \sum_{1\, \leq \, k_{1}}(\tanh (\beta
H))^{2k_{1}(p + 1)}  (\# (G(M);p + 1))^{- k_{1}}
\sum_{\mu_{1i}^{p + 1} (M):\, \, (\ref{4.32})}
 $$
\begin{equation}
\label{4.55}
||\sum_{i\, =\, 1}^{k_{1}} \mu_{1i}^{p + 1}
(M)||_{J,G(M)} \left( 1 + \alpha_{2}^{(N)} \left( \sum_{i_{1}\, =\, 1}^{k_{1}} \partial
\mu_{1i_{1}}^{p + 1} (M) ,\sum_{i_{1}\, =\, 1}^{k_{1}} \mu_{1i_{1}}^{p + 1} (M);
G(M)\right) \right)^{- 1}.
\end{equation}
By making use of the proof of the equality (\ref{4.55}) we get
$$
\lim_{G(M) \rightarrow {\bf Z}^{\times d}} ((k_{1})!)^{- 1} (\tanh
(\beta H))^{2k_{1}(p + 1)}(\# (G(M);p + 1))^{- k_{1}}
$$
$$
\times \left( \sum_{\chi^{p} \, \in \, B_{p}(P(G(M)),{\bf Z}_{2}^{add}), \, \,
|\chi^{p}|_{P(G(M))} \, =\, 2p + 2} \alpha (\chi^{p}; G(M))
\right)^{k_{1}} = \lim_{N \rightarrow \infty}
\lim_{G(M)\rightarrow {\bf Z}^{\times d}}
$$
$$
(\tanh (\beta H))^{2k_{1}(p + 1)} (\# (G(M);p + 1))^{- k_{1}} \sum_{
\mu_{1i_{1}}^{p + 1} (M):\, \, (\ref{4.32})} ||\sum_{i_{1}\, =\, 1}^{k_{1}}
\mu_{1i_{1}}^{p + 1} (M)||_{J,G(M)}
$$
\begin{equation}
\label{4.56}
\times \left( 1 + \alpha_{2}^{(N)} \left( \sum_{i_{1}\, =\, 1}^{k_{1}} \partial
\mu_{1i_{1}}^{p + 1} (M) ,\sum_{i_{1}\, =\, 1}^{k} \mu_{1i_{1}}^{p + 1} (M);
G(M)\right) \right)^{- 1},
\end{equation}
$k_{1} = 1,2,...$. All $(k_{1})!$ possible ordering of the different cochains
$\mu_{1i_{1}}^{p + 1} \in C^{p + 1}(P(G(M)),{\bf Z}_{2}^{add})$, $i = 1,...,k_{1}$,
give the same sum
$$
\sum_{i_{1}\, =\, 1}^{k_{1}} \mu_{1i_{1}}^{p + 1}
\in C^{p + 1}(P(G(M)),{\bf Z}_{2}^{add}).
$$
It explains the multiplier $((k_{1})!)^{- 1}$ in the left-hand side of the
equality (\ref{4.56}). By making use of the equality (\ref{4.55}) and summing
up the equalities (\ref{4.56}) we get
\begin{eqnarray}
\label{4.57}  \lim_{G(M) \rightarrow {\bf Z}^{\times d}} S_{p}(J,(\#
(G(M);p + 1))^{- 1/(2p\, + 2)} \beta^{- 1} \tanh (\beta H);G(M)) =
\nonumber \\ \lim_{G(M) \rightarrow {\bf Z}^{\times d}} \exp \Biggr\{
(\tanh (\beta H))^{2p + 2} (\# (G(M);p + 1))^{- 1} \times \nonumber \\
\sum_{\chi^{p} \, \in \, B_{p}(P(G(M)),{\bf Z}_{2}^{add}), \, \,
|\chi^{p}|_{P(G(M))} \, =\, 2p + 2} \alpha (\chi^{p}; G(M))
\Biggl\}.
\end{eqnarray}
By making use of the equality (\ref{4.24}) and of the proof of the
equality (\ref{4.57}) we can prove
$$
\lim_{G(M) \rightarrow {\bf Z}^{\times d}}
\frac{\partial}{\partial x} \left( \ln \frac{Z_{p}(J,(\# (G(M);p + 1))^{-
1/(2p\, + 2)} \beta^{- 1} x;G(M)) }{Z_{p}(0,(\# (G(M);p + 1))^{-
1/(2p\, + 2)} \beta^{- 1} x;G(M))} \right)_{x\, =\, \tanh
\beta H} =
$$
$$
2(p + 1)(\tanh \beta H)^{2p + 1} \times
$$
\begin{equation}
\label{4.58}
\lim_{G(M) \rightarrow {\bf Z}^{\times d}}
(\# (G(M);p + 1))^{- 1} \sum_{\chi^{p} \, \in \,
B_{p}(P(G(M)),{\bf Z}_{2}^{add}), \, \, |\chi^{p}|_{P(G(M))} \, =\,
2p + 2} \alpha (\chi^{p}; G(M)).
\end{equation}
The equality (\ref{4.58}) is proved for $p = 0$, $d = 1,2,3$ and for $p = 1$, $d = 2,3$.
The equality (\ref{4.58}) for $p = 0$, $d = 1$ and the constant interaction energy
$J(s_{j}^{1})$ coincides with the equality (\ref{4.18}). The equality (\ref{4.58})
for $p = 0$, $d = 2$ is proved in the paper \cite{9}.

\end{document}